\documentclass[fleqn,usenatbib]{mnras}

\usepackage[normalem]{ulem}

\usepackage[T1]{fontenc}

\DeclareRobustCommand{\VAN}[3]{#2}
\let\VANthebibliography\thebibliography
\def\thebibliography{\DeclareRobustCommand{\VAN}[3]{##3}\VANthebibliography}

\usepackage{graphicx}	
\usepackage{amsmath}	
\usepackage{amssymb}	
\usepackage{todonotes}
\usepackage{microtype}
\usepackage{xcolor}
\usepackage{float}

\def\HI{\ion{H}{I}~}
\def\Hi{\ion{H}{I}}
\def\HII{\ion{H}{II}~}
\def\xHI{x_{\rm \ion{H}{I}}}

\def\xb{\bar{x}_{\rm \ion{H}{I}}}

\def\lya{Ly$\alpha$~}
\defcitealias{Lidz2007}{L07}	
\defcitealias{Xu2019}{X19}

\usepackage{lipsum} 
\usepackage{comment}

\usepackage{tikz,xcolor,hyperref}
\definecolor{lime}{HTML}{A6CE39}
\DeclareRobustCommand{\orcidicon}{%
	\begin{tikzpicture}
	\draw[lime, fill=lime] (0,0) 
	circle [radius=0.16] 
	node[white] {{\fontfamily{qag}\selectfont \tiny ID}};
	\draw[white, fill=white] (-0.0625,0.095) 
	circle [radius=0.007];
	\end{tikzpicture}
	\hspace{-2mm}
}

\foreach \x in {A, ..., Z}{%
	\expandafter\xdef\csname orcid\x\endcsname{\noexpand\href{https://orcid.org/\csname orcidauthor\x\endcsname}{\noexpand\orcidicon}}
}

\usepackage{newtxtext,newtxmath}

\title[ large-scale 21-cm power spectrum]{The large-scale 21-cm power spectrum from reionization}

\author[I. Georgiev et al.]{
Ivelin Georgiev $^{1}$\thanks{E-mail: ivelin.georgiev@astro.su.se}\orcidA,
Garrelt Mellema$^{1}$\orcidB, Sambit K. Giri$^{2}$\orcidC and Rajesh Mondal$^{1}$\orcidD
\\
$^{1}$The Oskar Klein Centre, Department of Astronomy, Stockholm University, AlbaNova, SE-10691 Stockholm, Sweden\\
$^{2}$Institute for Computational Science, University of Zurich,
Winterthurerstrasse 190, 8057 Zurich, Switzerland\\
}

\date{Accepted XXX. Received YYY; in original form ZZZ}

\pubyear{2021}

\begin{document}
\label{firstpage}
\pagerange{\pageref{firstpage}--\pageref{lastpage}}
\maketitle

\begin{abstract}
Radio interferometers, such as the Low-Frequency Array and the future Square Kilometre Array, are attempting to measure the spherically averaged 21-cm power spectrum from the Epoch of Reionization. Understanding of the dominant physical processes which influence the power spectrum at each length-scale is therefore crucial for interpreting any future detection. We study a decomposition of the 21-cm power spectrum and quantify the evolution of its constituent terms for a set of numerical and semi-numerical simulations of a volume of $(714~\mathrm{Mpc})^3$, focusing on large scales with $k\lesssim 0.3$~Mpc$^{-1}$. We find that after $\sim 10$ per cent of the Universe has been ionized, the 21-cm power spectrum follows the power spectrum of neutral hydrogen fluctuations, which itself beyond a certain scale follows the matter power spectrum. Hence the signal has a two-regime form where the large-scale signal is a biased version of the cosmological density field, and the small-scale power spectrum is determined by the astrophysics of reionization. We construct a bias parameter to investigate the relation between the large-scale 21-cm signal and the cosmological density field. We find that the transition scale between the scale-independent and scale-dependent bias regimes is directly related to the value of the mean free path of ionizing photons ($\lambda_{\mathrm{MFP}}$), and is characterised by the empirical formula $k_{\mathrm{trans}} \approx 2/\lambda_{\mathrm{MFP}}$. Furthermore, we show that the numerical implementation of the mean free path effect has a significant impact on the shape of this transition. Most notably, the transition is more gradual if the mean free path effect is implemented as an absorption process rather than as a barrier.

\end{abstract}




\section{Introduction}

During the Dark Ages, the neutral hydrogen atoms (\Hi) that formed in the epoch of recombination traced the structure of the Universe and permeated the intergalactic medium (IGM). Due to gravitational interactions, these baryons gathered in the dark matter haloes forming along the cosmic web. This process led to the formation of the first stars and the early galaxies they occupied, and the ionizing radiation from these is thought to have initiated the Epoch of Reionization (EoR). During this epoch, the Universe experienced its last known phase transition, from neutral to ionized, as UV radiation interacted with the neutral hydrogen atoms \citep[see e.g. ][for reviews]{Morales2010, 2019ConPh..60..145W}. 

Observations of this process contain a wealth of information about the first luminous sources and is currently an active area of research. Measurements of the Thomson scattering optical depth of the Cosmic Microwave Background (CMB) \citep{Planck2020} indicate $\tau =0.051 \pm 0.006$, suggesting that the midpoint of EoR lies around a redshift of $z \sim 8$ (for an instantaneous reionization model). Additionally, the detection of the Gunn-Peterson effect \citep{Gunn-Peterson1965} in high-redshift quasar spectra provides estimates of the $\Hi$ fraction at high-$z$, as it removes the signal by scattering Lyman-$\alpha$ (\lya) photons \citep{Fan2006,Mortlock2011,Greig2017,Banados2018,Greig2019,Wang2020,Yang2020,Becker2021}. Furthermore, the drop of the \lya luminosity function for $z \geq 7$, due to the decrease of number of \lya emitters, suggests the presence of a neutral hydrogen rich Universe with only small ionized bubbles \citep{Konno2014}.

Neutral hydrogen itself is also one of the most promising observational probes of the first billion years of the Universe. Its hyperfine structure causes the emission or absorption of 21-cm radiation. The resulting 21-cm signal is indicative of \Hi~ present in the IGM, and will diminish as ionizing radiation escapes into the IGM during the EoR. One method of detection involves measuring the sky-averaged 21-cm signal using single-dish antenae such as SARAS \citep{Patra2015}, SARAS2 \citep{Singh2018}, LEDA \citep{Price2018}, EDGES \citep{Bowman2010}, and EDGES 2 \citep{Monsalve2017}. The EDGES team have claimed a detection at 78 MHz \citep{Bowman2018}, but the distinct and deep absorption profile has sparked a range of investigations, most recent of which is the {\sc SARAS3} non-detection \citep[see][ which also provides an overview of previous work]{Singh2022}.

Meanwhile, low-frequency radio interferometers offer the prospect of detecting the fluctuations in the 21-cm signal. The Precision Array for Probing the Epoch of Reionization (PAPER)\footnote{\url{http://eor.berkeley.edu}} \citep{Parsons2010}, the Low-Frequency Array (LOFAR)\footnote{\url{www.lofar.org}} \citep{vanHaarlem2013}, the Hydrogen Epoch of Reionization Array (HERA)\footnote{\url{https://reionization.org}} \citep{DeBoer2017}, the Murchison Widefield Array (MWA)\footnote{\url{www.mwatelescope.org}} \citep{Bowman2013}, and the prospective Square Kilometre Array (SKA)\footnote{\url{www.skatelescope.org}} \citep{2015aska.confE...1K} are telescopes of this type. One possible statistic, which is an essential stepping stone for understanding the nature of the EoR, is the power spectrum of the 21-cm signal \citep[][hereafter L07]{Furlanetto2006,Lidz2007}. Several of the systems mentioned have been reporting upper limit values on the power spectra at various redshifts and scales; GMRT \citep{Paciga2013}, PAPER \citep{Parsons2014}, LOFAR \citep{Mertens2020}, HERA \citep{HERA2021}, MWA \citep{Trott2020,Yoshiura2021}. These interferometers typically have the largest sensitivity for the power spectra at large scales or low $k$-modes, due to the large numbers of short baselines they contain. Whilst a detection has not yet been made, the upper limits have been employed in order to exclude non-viable models and provide constraints on the EoR \citep{ghara2020constraining,Mondal2020,Greig2021,Ghara2021,HERA2021a}. These measurements serve as a precursor to SKA, which is designed to have an increased sensitivity and should be able to produce tomographic imaging data \citep{Mellema2015}. 

There are therefore good prospects for studying the large-scale 21-cm power spectrum from reionization. These scales are attractive to study as they should be simpler to understand since many of the small scale astrophysical effects will be averaged over. It is also commonly thought that the largest scales will follow the matter power spectrum, thus providing a cosmological measurement \citep{Xu2019}.

The aims of this paper are to explore the behaviour of the large-scale 21-cm power spectrum, its evolution throughout reionization, the connection with the sizes of ionized regions and the matter power spectrum, as well as how all this depends on the astrophysical parameters of reionization. 
In the context of our study we will consider modes with $k\lesssim 0.3$~Mpc$^{-1}$ to constitute large scales.

We base our study on the results of both numerical and semi-numerical simulations of reionization and will use the decomposition of the 21-cm power spectrum into its components, as introduced by \citetalias{Lidz2007}. Due to the size of their simulation volume ($<100$~Mpc per side), these authors could only focus on smaller scales and by using a $(714~\mathrm{Mpc})^3$ larger volume we extend their study to large scales. In order to elucidate the relation between the 21-cm and matter power spectra, we also explore the scale-dependent bias with respect to the cosmological matter power spectrum, similar to what was done in \cite{Xu2019}.

The format of the paper is as follows. We establish the relevant theoretical framework in Sec.~\ref{sec:Theory} and the simulations employed in the analysis in Sec.~\ref{Sec:Simulations}. In Sec.~\ref{Sec:DecStudy} we proceed by decomposing the signal investigating which terms dominate during the various stages of reionization. We highlight the importance of the mean free path (MFP) of ionizing photons has on the large-scale power spectra and examine its effect in Sec.~\ref{Sec:MFP_Study}. We present a conclusion and summary of our results in Sec.~\ref{Sec:Conclusions}. An appendix with additional information about methods to model the MFP effect in reionization simulations is included in Appendix \ref{appendix:MFP_application}. Throughout the paper we adopt a flat $\Lambda$CDM cosmology with parameters
($\Omega_{\mathrm{m}},\Omega_{\mathrm{b}},h,n_{s},\sigma_{8}$) = (0.27, 0.044, 0.7, 0.96, 0.8) consistent with \textit{WMAP} \citep{hinshaw13} and \textit{Planck} results \citep{planck14}.

\section{Theory}
\label{sec:Theory} 

The 21-cm line is the result of the spin-flip transition, occurring in the ground state of neutral hydrogen. The 21-cm radiation obsereved by radio interferometry telescopes is defined as the differential surface brightness temperature against the CMB. In the Rayleigh-Jeans limit it can be written as follows \citep{Mellema2013}
\begin{equation}
\label{eq::Tb}
\delta T_{\mathrm{b}} (\mathbf{r}, z) =\frac{1-e^{-\tau(\mathbf{r}, z)}}{1+z}(T_{s}(\mathbf{r}, z)-T_{\mathrm{CMB}}(z))\,,
\end{equation}
where $\tau$ is the optical depth of the gas and $T_{s}$ is the spin temperature. 
The spin temperature characterizes the level populations of the 21-cm transition and has to be different from $T_\mathrm{CMB}$ to produce an observable signal. At the redshifts of reionization $T_{s}$ is expected to be similar to the kinetic temperature of the gas due to the frequent interaction of the intergalactic {\sc H~I} with \lya photons produced by star forming galaxies \citep{Field1959, Wouthuysen1952}. X-ray heating caused by the same galaxies is expected to increase this kinetic temperature \citep{Pritchard2007}. 
The scales of interaction leading up to the 21-cm signal are thus well-tuned to exploring the influence of the first sources on the state of IGM. 

Assuming an optically thin gas $\tau \ll 1$ at high redshifts, Eq.~\ref{eq::Tb} can be rewritten as 
\begin{align}
\label{eq::TbS}
\delta T_{\mathrm{b}} (\mathbf{r}, z) = T_{0} (\mathbf{r},z) \xb(1+\delta_{\HI}(\mathbf{r}))(1+\delta_{\rho}(\mathbf{r})) \ , 
\end{align}
\begin{align}
&T_0 (\mathbf{r},z) \approx 27 
\bigg(\frac{1+z}{10}\bigg)^{1/2} \bigg(\frac{T_{\mathrm{s}}(\mathbf{r},z) -T_{\mathrm{CMB}}(z) }{T_{\mathrm{s}}(\mathbf{r},z) }\bigg) 
\nonumber
\\ &\bigg( \frac{\Omega_{\mathrm{b}}}{0.044}  \frac{h}{0.7} \bigg) \bigg( \frac{\Omega_{\mathrm{m}}}{0.27}  \bigg)^{-1/2} 
\bigg( \frac{1-Y_{\mathrm{p}}}{1-0.248} \bigg)
~\mathrm{mK},  
\end{align}
Here the variable $\rho$ is the mass density, whilst $\delta_{\rho}= \rho/\bar{\rho} -1 $ is the corresponding fluctuation. Similarly, $\delta_{\HI}$ denotes the fluctuation of the neutral fraction $\xHI$ field. The term $T_{0}$ collects all the cosmological parameters as well as the spin temperature dependence. Note that within this work we do not consider the effects of redshift space distortions.
We analyse the 21-cm power spectra of the field in Eq.~\ref{eq::TbS}. We also define the dimensionless cross power spectrum between a field $a$ and $b$ at a certain wavenumber $k$ as $\Delta^{2}_{\mathrm{a,b}}(k)=k^{3} P_{\mathrm{a,b}}(k)/(2\pi^{2})$.
In the limit where $T_{\mathrm{s}}\gg T_{\mathrm{CMB}}$ the 21-cm dimensionless power spectrum is derived by linearising $\delta T_{\mathrm{b}} (\mathbf{r}, z)$ \citepalias[see e.g.][]{Lidz2007}

\begin{align}
\label{Eq::Decomposition}
\Delta_{\mathrm{21cm}}^{2} (k) = T_{0}^{2} \xb^{2}\Big( \Delta_{\delta_{\rho},\delta_{\rho}}^{2}(k)+\Delta_{\delta_{\HI},\delta_{\HI}}^{2}(k) \\ +2\Delta_{\delta_{\rho},\delta_{\HI}}^{2} (k)
+2\Delta_{\delta_{\rho}\delta_{\HI},\delta_{\HI}}^{2}(k) +2\Delta_{\delta_{\rho}\delta_{\HI},\delta_{\rho}}^{2}(k) \nonumber \\
+\Delta_{\delta_{\rho}\delta_{\HI},\delta_{\rho}\delta_{\HI}}^{2} (k) \Big) \nonumber.
\end{align}
As we seek to study the decomposition terms, we redefine the 21-cm dimensionless power spectrum as
\begin{equation}
    \Delta_{21 \star}^{2} (k) = \frac{ \Delta_{\mathrm{21cm}}^{2} (k)}{T_{0}^{2} \xb^{2}},
\end{equation}
where $\delta_{21 \star} = (1+\delta_{\rho}) (1+\delta_{\HI})$. The variable $\rho$ is the density, whilst $\delta_{\rho}= \rho/\bar{\rho} -1 $ is the corresponding fluctuation. Similarly, $\delta_{\HI}$ denotes the fluctuation of the neutral fraction $\xHI$ field.

In the following sections we refer to this decomposition and the role of each of its constituting terms. First-order terms are defined as the auto-correlation power spectra and the cross-power spectrum between the two perturbation fields.
We define them as follows
\begin{itemize}
    \item $\Delta_{\delta_{\rho},\delta_{\rho}}^{2}$ is the density term,
    \item $\Delta_{\delta_{\HI},\delta_{\HI}}^{2}$ is the neutral hydrogen term,
    \item $\Delta_{\delta_{\rho},\delta_{\HI}}^{2}$ is the cross term between density and neutral hydrogen,
    \item $\Delta_{\delta_{\rho}\delta_{\HI},\delta_{\HI}}^{2}$, $\Delta_{\delta_{\rho}\delta_{\HI},\delta_{\rho}}^{2}$ are three-point corrections and referred to as higher-order terms,
    \item $\Delta_{\delta_{\rho}\delta_{\HI},\delta_{\rho}\delta_{\HI}}^{2}$ is the auto-correlation four-point term.
\end{itemize}
It has often been assumed that the higher-order terms have a negligible contribution. However, \citetalias{Lidz2007} highlighted the importance of the higher-order terms at small-scales. They also inferred that the largest scales should also be subject to higher-order corrections and that simulations of box size larger than 100~Mpc are required to study this effect. We employ a suite of such large box simulations, described in Sec.~\ref{Sec:Simulations}, to further this analysis and present the results in Sec.~\ref{Sec:DecStudy}. Within this work all length scales are given in comoving units, typically Mpc.

\subsection{Bias Parameter}
Bias models are a well known tool for quantifying the dark matter density distribution in galaxy surveys. This is a potential avenue of investigation for the cosmological 21-cm signal and several papers have studied this possibility with interesting findings \citep{Santos2006,McQuinn2018,Hoffmann2019}. 

In order to test how the largest scales are affected by the neutral hydrogen fluctuations and their relation to the cosmological matter power spectrum, we construct a bias of the form 
\begin{equation}
    \delta_{21\star} (k) = b(k) \delta_{\rho}(k),
\end{equation}
which can be expressed as
\begin{equation}
b^{2}(k) = \frac{\Delta_{21 \star}^{2} (k)}{\Delta_{\delta_{\rho},\delta_{\rho}}^{2}(k)}.\\
\end{equation}

Extracting the bias would involve detecting the 21-cm signal (numerator) and dividing against the density term (denominator), determined by the chosen cosmology.
However, in this form, the large-scale bias is prone to sampling errors such as shot noise and stochasticity. We resolve this by extracting the bias parameter by cross- correlating the $\delta_{21 \star}$ and $\delta_{\rho}$ fields, see \citet{Seljak2009} and eq. (9) of \citet{Xu2019}.
The bias is then written as
\begin{equation}
\label{eq:Bias}
b(k) = \frac{\Delta_{21 \star,\delta_{\rho}}^{2} (k)}{\Delta_{\delta_{\rho},\delta_{\rho}}^{2}(k)},\\
\end{equation}
where we can express the $\Delta_{21 \star,\delta_{\rho}}^{2} (k)$ term through the decomposition
\begin{equation}
\Delta_{21 \star,\delta_{\rho}}^{2} (k) = \Delta_{\delta_{\rho},\delta_{\HI}}^{2}(k)+ \Delta_{\delta_{\rho},\delta_{\rho}}^{2}(k)+\Delta_{\delta_{\rho}\delta_{\HI},\delta_{\rho}}^{2}(k).
\end{equation}

On scales where $b(k)$ does not change with $k$ the 21-cm signal can be described as a biased version of the density power spectrum. In these regimes the 21-cm power spectrum can then be said to be a probe of cosmology. One of our goals is to explore for which periods and scales such a scale-independent bias exists.

\section{Simulations}
\label{Sec:Simulations}
The process of reionization is complex in nature and difficult to describe analytically, as it involves physical and stochastic interactions of varying range and duration. In order to model the signal accurately, state-of-the-art simulations are needed to bridge the gap between astrophysics and cosmology. On one hand, small-scale simulations are necessary to predict the structure of the first galaxies and the stellar objects that habitate them. On the other hand, large-scale simulations are needed to understand the large-scale IGM and source distribution \citep{Iliev2014,Kaur2020}.
Hence, a hybrid approach is chosen where small-scale physics is included through subgrid recipes in simulations of several hundreds of Mpc \citep[e.g.][]{2016MNRAS.456.3011D}. 

Simulations also provide a virtual laboratory to study the observables of reionization. Prior to any detection, the goal of such examinations is to establish an intuition of the key parameters which influence the probes. Within this body of work we choose to study the impact of the ionization efficiency, minimum dark matter halo mass containing sources and mean free path of ionizing photons on the 21-cm power spectrum.

\subsection{{\sc 21cmFAST} simulations}
\label{sec:21cmfast}
\begin{figure}
\includegraphics[width=\columnwidth]{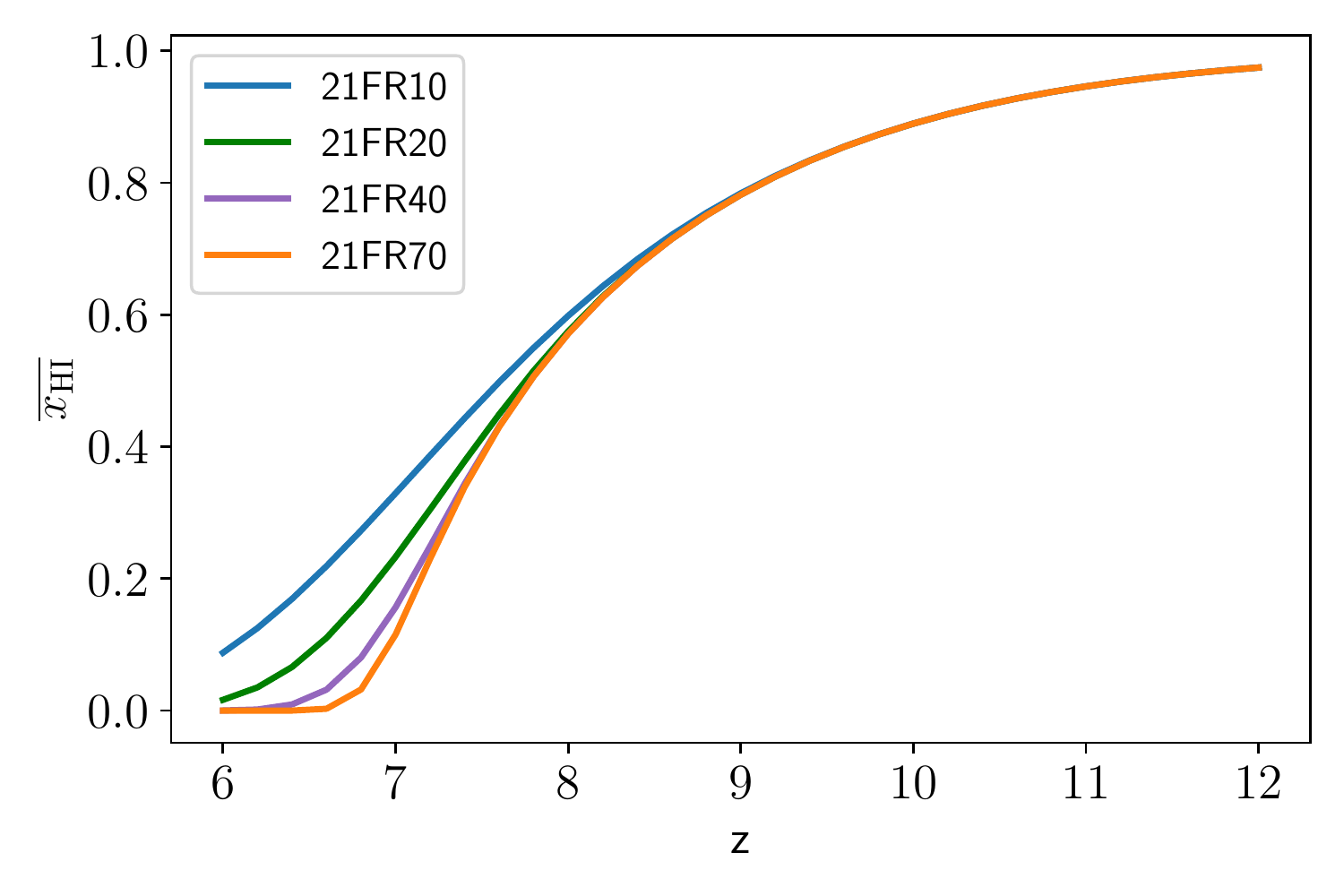}
\includegraphics[width=\columnwidth]{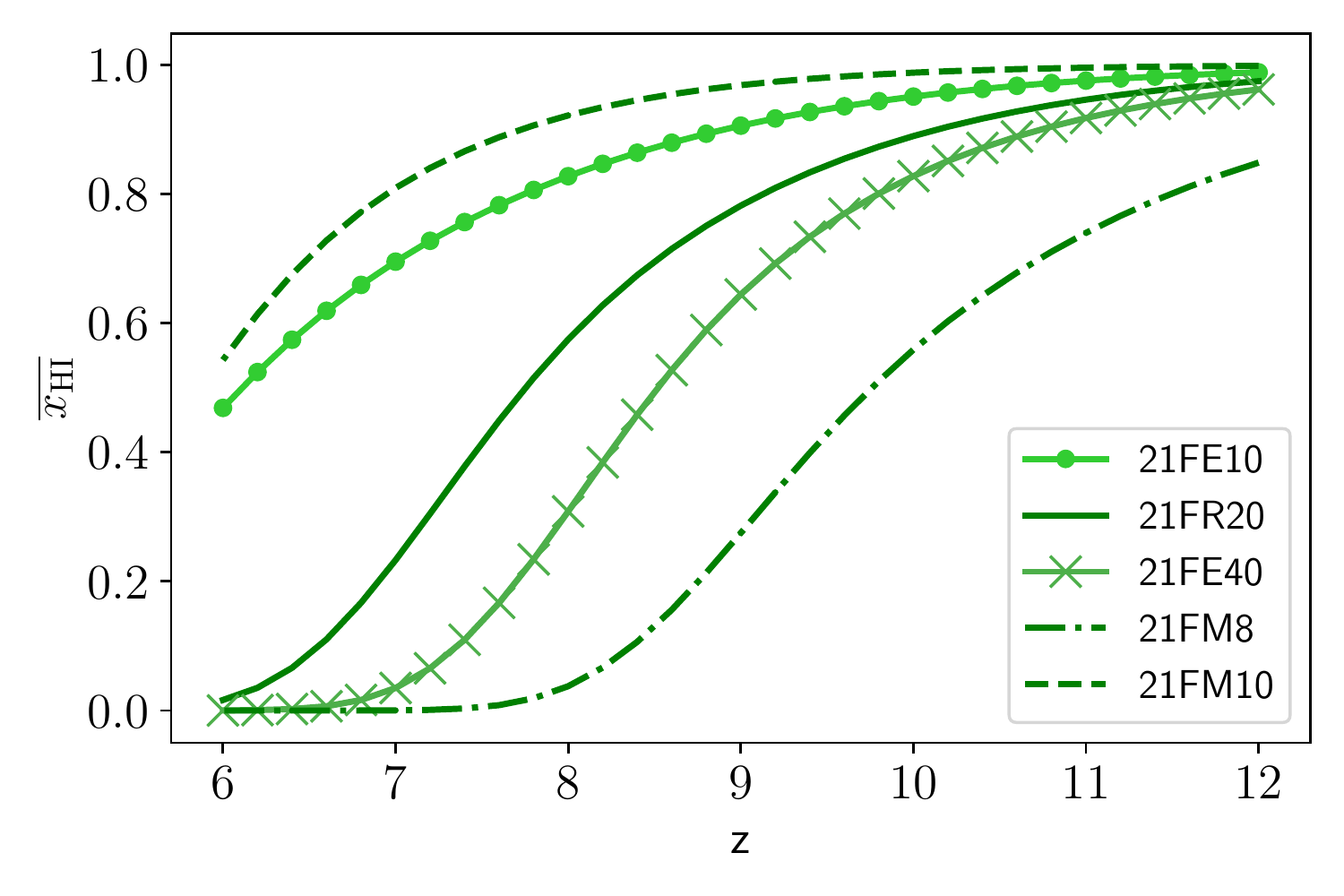}
\caption{Top Panel: Reionization histories of four {\sc 21cmFAST} models which only differ in the value of the mean free path parameter $R_{\mathrm{max}}$. All other parameters are identical ($\zeta = 0.25$, $M_{\mathrm{min}} = 10^{9} M_{\odot}$). Bottom Panel: Likewise for a set of {\sc 21cmFAST} models in which either the minimal halo mass  (``M'' models) or the efficiency (``E'' models) differs from the fiducial model 21FR20. Parameters for all models are listed in Tab. \ref{tab:21cmSimulations}.}
\label{fig:21cmFAST_Histories}
\end{figure}

We run a suite of semi-numerical simulations using the publicly available code {\sc 21cmFAST} v1.2 \citep{Mesinger2007, Mesinger2011}. {\sc 21cmFAST} provides a fast and effective method of examining the 21-cm signal and parameter estimation. 
In this code, the density field is simulated with the Zel'dovich approximation while reionization is simulated using the excursion set approach introduced in \citet{Furlanetto2004}.

We produce four semi-numerical reionization models of which the fiducial source parameters are the ionizing efficiency $\zeta = 25$ and the minimum halo mass $M_{\mathrm{min}} = 10^{9} M_{\odot}$. The choice of mean free path parameter $R_{\mathrm{max}}$ covers the range 10, 20, 40 and 70~Mpc\footnote{We are thus using a constant and uniform $R_{\mathrm{max}}$ value. Newer versions of { \sc 21cmFAST} include more physically motivated implementations of the MFP effect in which it becomes both time and position dependent \citep[see][]{Sobacchi2014}.}. We choose the model with $R_{\mathrm{max}} = 20~\mathrm{Mpc}$ as the fiducial one and produce four additional simulations, where we vary the $\zeta$ and $M_{\mathrm{min}}$ parameters. The reionization history of these eight models are presented in Fig.~\ref{fig:21cmFAST_Histories} with information on their source parameters given in Tab. \ref{tab:21cmSimulations}. We choose the size of the simulations to have a side length of $L_{\mathrm{box}} \approx 714 $~Mpc, all models are then processed through a grid size $300^{3}$.
We will refer to these simulations by the labels given in Tab. \ref{tab:21cmSimulations}, as 21FR10, 21FR20, etc.

\begin{table}
\centering
\caption{{\sc 21cmFAST} simulation source parameters. } 
\label{tab:21cmSimulations}
\begin{tabular}{c|c|c|c} 

Label & $\zeta$ & \multicolumn{1}{p{1cm}|}{$M_{\mathrm{min}} \newline [M_{\odot}]$} & \multicolumn{1}{p{1cm}|}{$R_{\mathrm{max}} \newline $[Mpc]} \\ \hline
\multicolumn{1}{p{2cm}|}{21FR10, \textbf{21FR20}, 21FR40 21FR70
}  & 25 & $10^{9}$ & \multicolumn{1}{p{1cm}|}{10, 20, \newline 40, 70} \\
\\ \hline
\multicolumn{1}{p{2cm}|}{21FE10, \textbf{21FR20}, \newline 21FE40}  & \multicolumn{1}{p{1cm}|}{10, 25, \newline 40} & $10^{9}$ & 20 \\
\\ \hline
\multicolumn{1}{p{2cm}|}{21FM8, \textbf{21FR20}, \newline 21FM10}  & 25 & \multicolumn{1}{p{1cm}|}{$10^{8},10^{9}$, \newline $10^{10}$} & 20 \\
\end{tabular}

\end{table}

\subsection{{\sc\texorpdfstring{$\mathrm{C}^{2}$}-Ray} simulations}

\begin{table}
\centering
\caption{{\sc C$^2$-Ray} simulation source parameters. } 
\label{tab:C2RaySimulations}
\begin{tabular}{c|c|c|c|c} 

Label & \multicolumn{1}{p{1cm}|}{Source \newline Model}  & \multicolumn{1}{p{1cm}|}{$M_{\mathrm{min}}\newline [M_{\odot}]$} & \multicolumn{1}{p{1cm}|}{MFP \newline Method} & \multicolumn{1}{p{1cm}|}{$\lambda_{\mathrm{MFP}} \newline $[Mpc]} \\ \hline
C2RMAX10 & $\zeta=50$ & $10^{9}$ & spherical barrier& 10\\
\\ \hline
C2LLS10  & $\zeta=50$ & $10^{9}$ & absorption & 10\\

\end{tabular}
\end{table}

\begin{figure}
\includegraphics[width=\columnwidth]{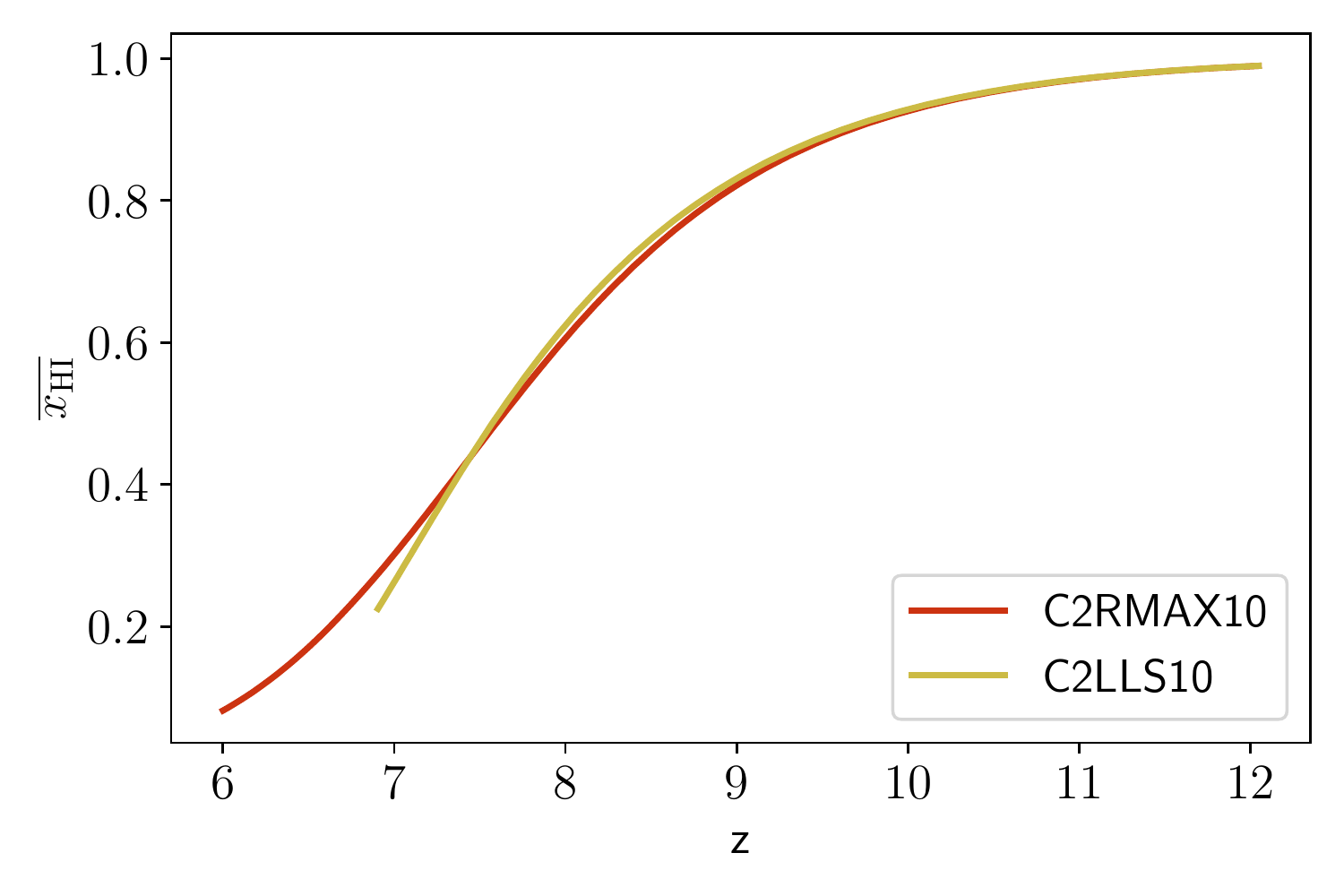}
\caption{Reionization histories for the two {\sc C$^2$-Ray} models which only differ in the implementation of the MFP effect. C2RMAX10 is highlighted in red, whilst C2LLS10 is presented in yellow. Parameters for all models are listed in Table \ref{tab:C2RaySimulations}. }
\label{fig:C2Ray_History}
\end{figure}

We briefly highlight the procedure of producing the fully numerical simulations studied in this paper. We follow the approach described in \citet{Mellema2006b} and \citet{iliev2006}. 
The density field and haloes are formed and tracked using the $N$-body simulation {\sc CUBEP$^{3}$M} code \citep{Harnois2013}. The resulting field is then post-processed using the {\sc $\mathrm{C}^2$Ray} radiative transfer code \citep{Mellema2006}, where haloes are assigned an ionizing luminosity.

We investigate a comoving volume of box length $L_{\mathrm{box}}\approx 714$~Mpc simulated with $\mathrm{CUBEP}^{3}$M using $N_{\mathrm{particles}} = 6912^{3}$ and a particle mass of $M_{\mathrm{particles}} = 4.05 \times 10^{7} M_{\odot}$. Haloes were identified on the fly through the spherical overdensity halo-finder method \citep{Watson2013}, using at least 25 particles thus yielding a minimum halo mass of about $10^9 M_{\odot}$. The radiative transfer was performed on gridded versions of the density fields obtained, using a mesh size of $300^{3}$ \citep[see][]{Giri2019b, Mondal2020a}. This simulation was run as part of the PRACE (Partnership for Advanced Computing in Europe) project {\sc PRACE4LOFAR}.

For the radiative transfer simulations the haloes were assigned an ionizing luminosity proportional to the mass of the haloes. In this normalisation, we first calculate the total rate of ionizing photons produced in the volume by multiplying the growth rate of haloes with a constant $\zeta$ and then distribute these over the individual haloes assuming a linear relation between halo mass and ionizing luminosity. This source model makes sure that the total number of ionizing photons produced up to a given redshift is proportional to the fraction of matter collapsed into dark matter haloes at that redshift. We refer to \citet{giri2021measuring} for more details. This source model is similar to the one used in the {\sc 21cmFAST} simulations described above.

The effect of a finite mean free path for ionizing photons is also included in these simulations. Slightly different implementations of this effect are used, which are described in more detail in Appendix~\ref{appendix:MFP_application}.

Table~\ref{tab:C2RaySimulations} lists the key parameters of the {\sc C$^2$-Ray} simulations used in this paper. The reionization histories for these simulations are shown in Fig.~\ref{fig:C2Ray_History}. 

\section{Study of large-scale decomposition terms}
\label{Sec:DecStudy}
In this section we study the evolution of the large-scale 21-cm power spectrum $\Delta_{\mathrm{21cm}}^{2}$ through its constituent terms, as defined in Eq.~\ref{Eq::Decomposition}. We choose the 21FR20 simulation (Tab. \ref{tab:21cmSimulations}) as our fiducial model. At the end of the section we compare against the results for a selection of other {\sc 21cmFAST} and {\sc C$^2$-Ray} simulations. We divide our analysis into subsections based on the evolution of the average neutral fraction of hydrogen. 

The 21-cm power spectrum and its decomposition terms from Eq.~\ref{Eq::Decomposition} are plotted against $k$ in Fig.~\ref{fig:21cmFast_DecTerms}. Results for $k > 0.6~\mathrm{Mpc}^{-1}$ are included for the sake of completeness, but may be prone to aliasing effects \cite[see sec. 6.4]{Mao2012} and are not considered within this body of work. Each panel corresponds to a different stage of reionization, relevant for the subsections below. Positive and negative values are represented by solid and dashed lines, respectively. 

In the upper panel of Fig.~\ref{fig:21cmFast_DecFrac}, we plot for $k = 0.1 ~\mathrm{Mpc}^{-1}$ the relative fraction of each decomposition term against the globally averaged neutral fraction $\bar{x}_\mathrm{HI}$. The sum of all of the plotted terms at each value of $\bar{x}_\mathrm{HI}$ adds up to one. This graph illustrates which decomposition terms are the dominant ones at different stages of reionization. We only show this for one $k$ value as the results do not differ substantially over the range $0.015~\mathrm{Mpc}^{-1}<k<0.3 ~\mathrm{Mpc}^{-1}$.

\subsection{Early reionization period \texorpdfstring{( $1.00>\xb>0.95$~)} )}
\label{subsec:Early_EoR}

\begin{figure*}

\includegraphics[width=.49\linewidth]{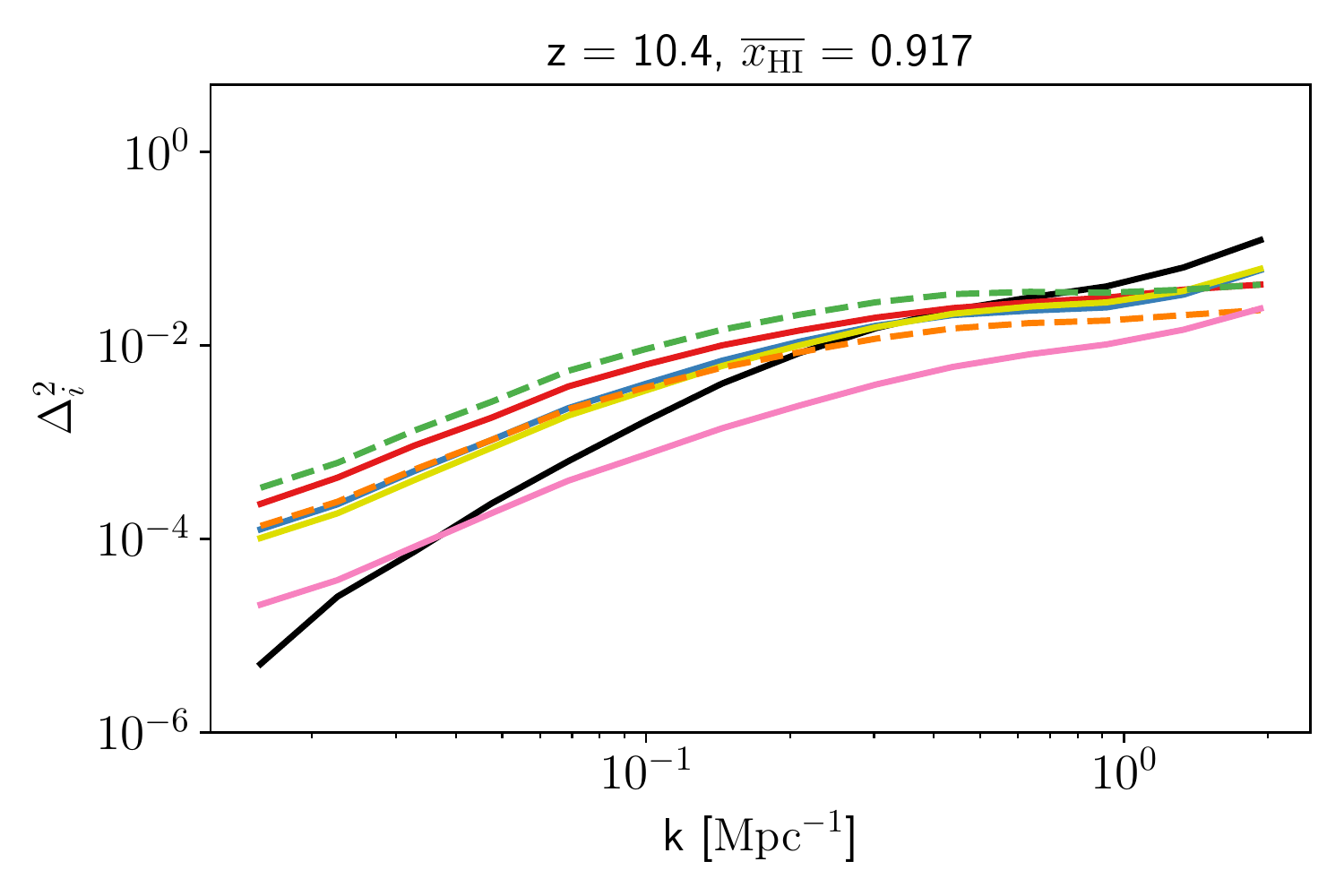}
\includegraphics[width=.49\linewidth]{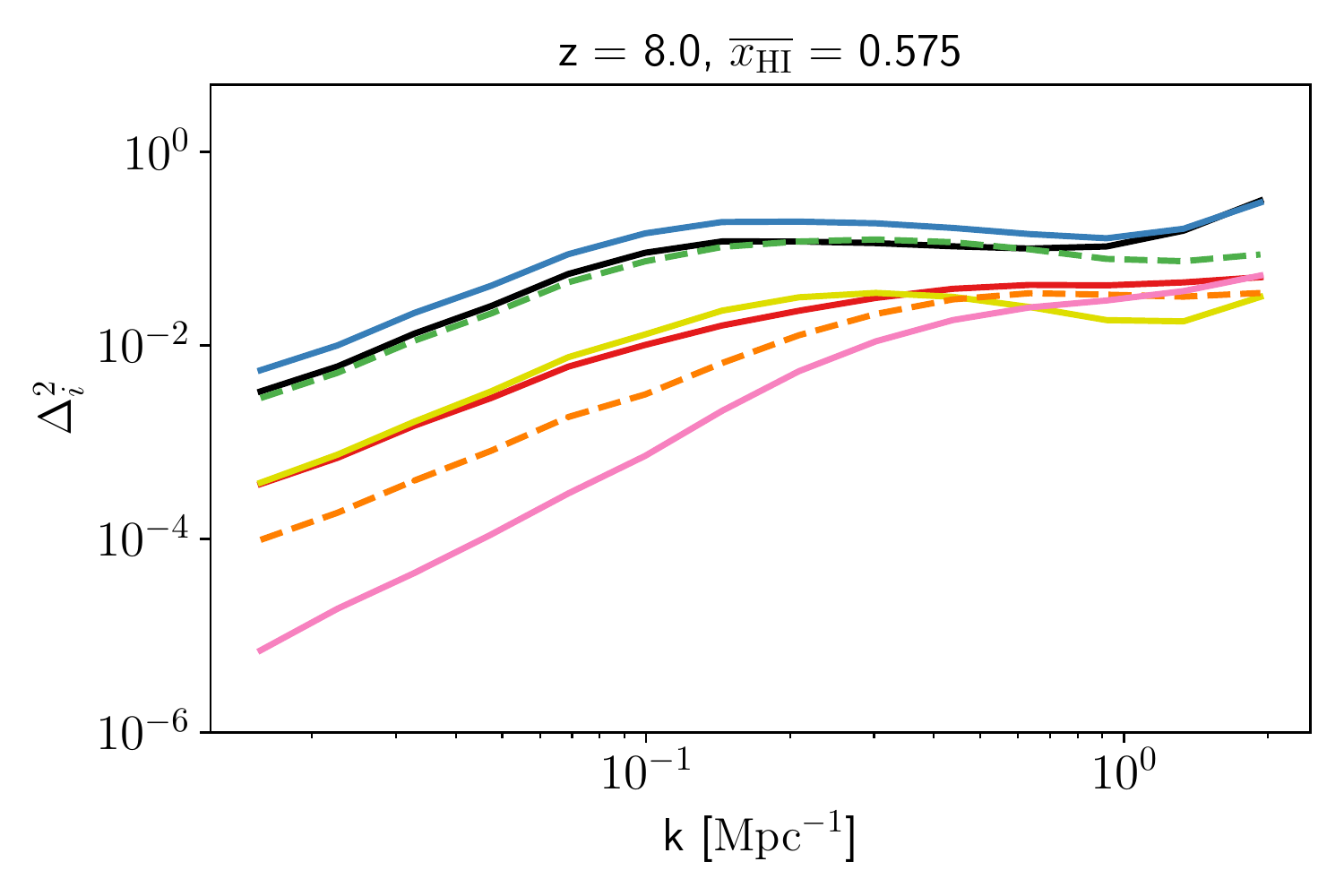}
\includegraphics[width=.49\linewidth]{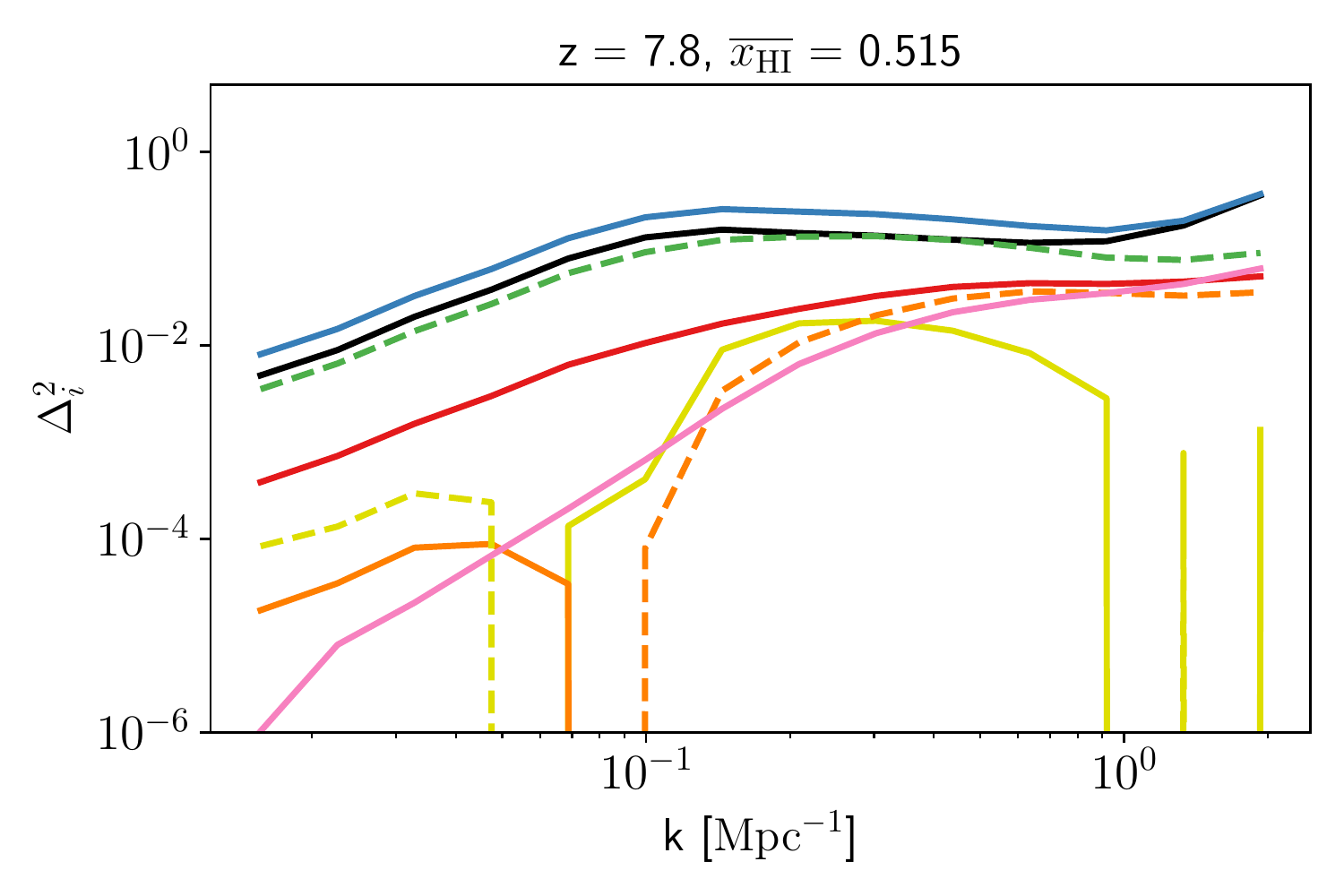}
\includegraphics[width=.49\linewidth]{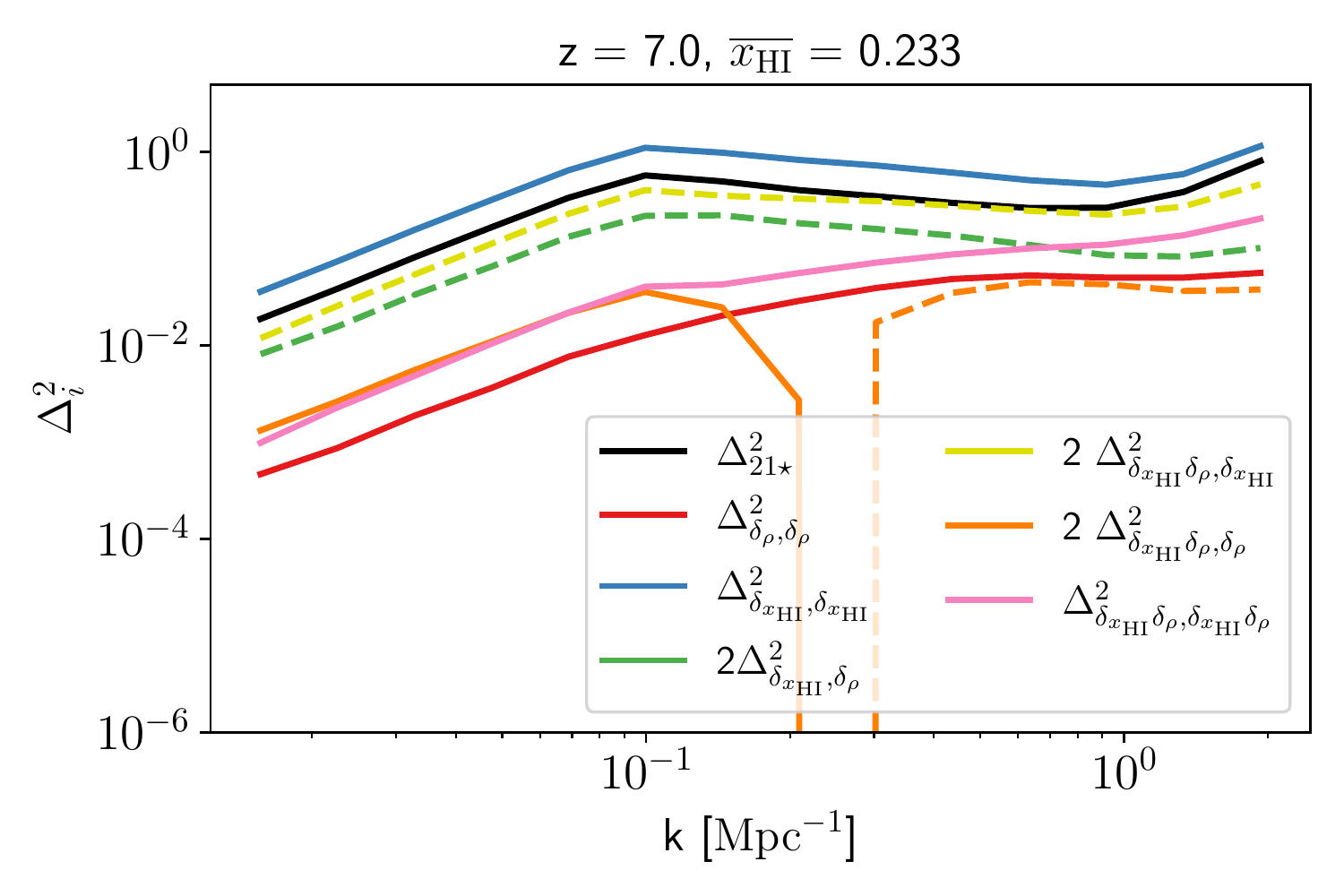}
\caption{Power spectra of the decomposition terms presented in Eq. ~(\ref{Eq::Decomposition}) for our fiducial simulation 21FR20 during different stages of reionization. Top left: equilibration period, Top right: before the mid-point period, Bottom left: mid-point period, Bottom right: later stages of reionization. Solid lines mark positive terms, dashed lines represent the absolute value of the negative terms.
}
\label{fig:21cmFast_DecTerms}
\end{figure*}

\begin{figure}
\includegraphics[width=\columnwidth]{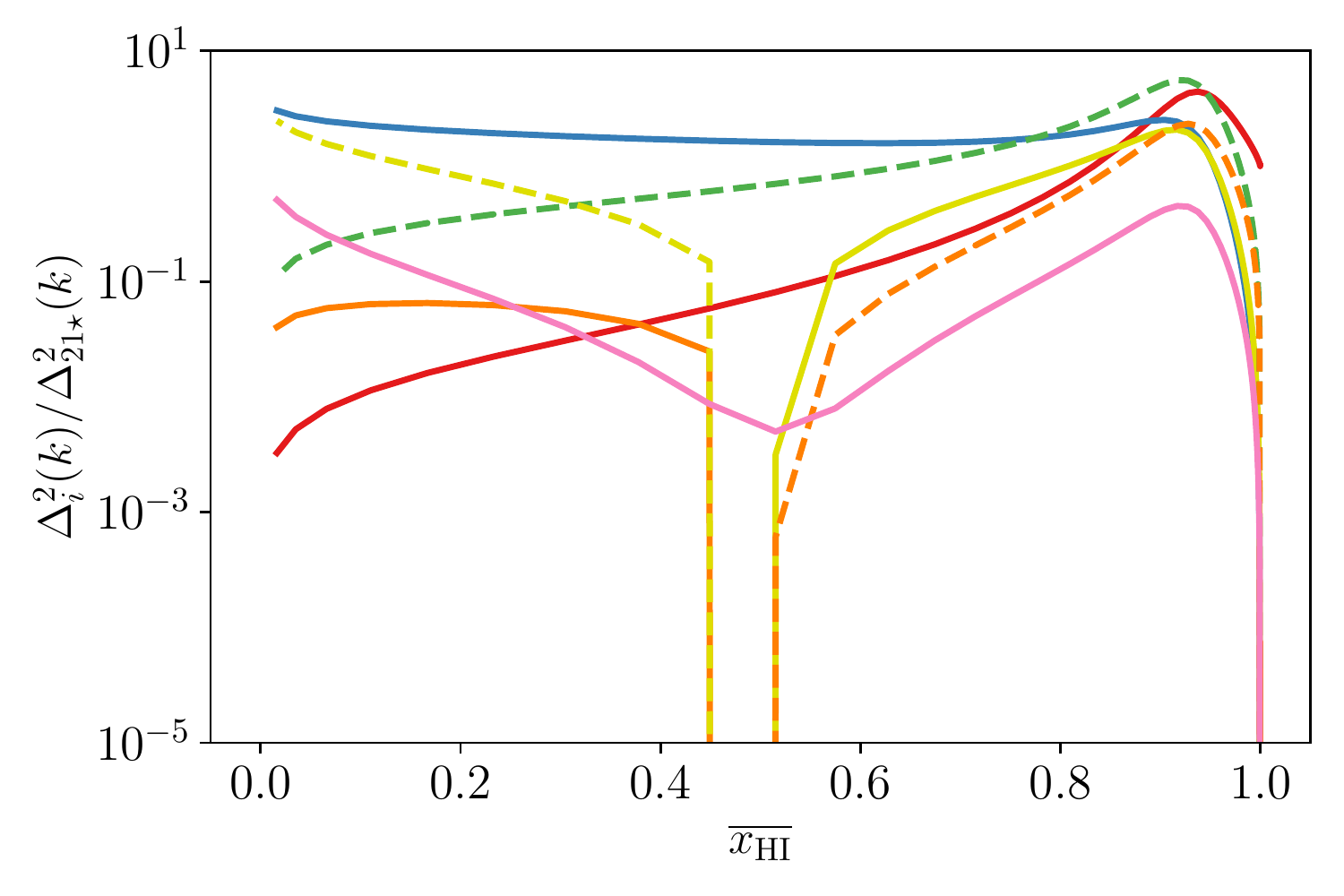}
\includegraphics[width=\columnwidth]{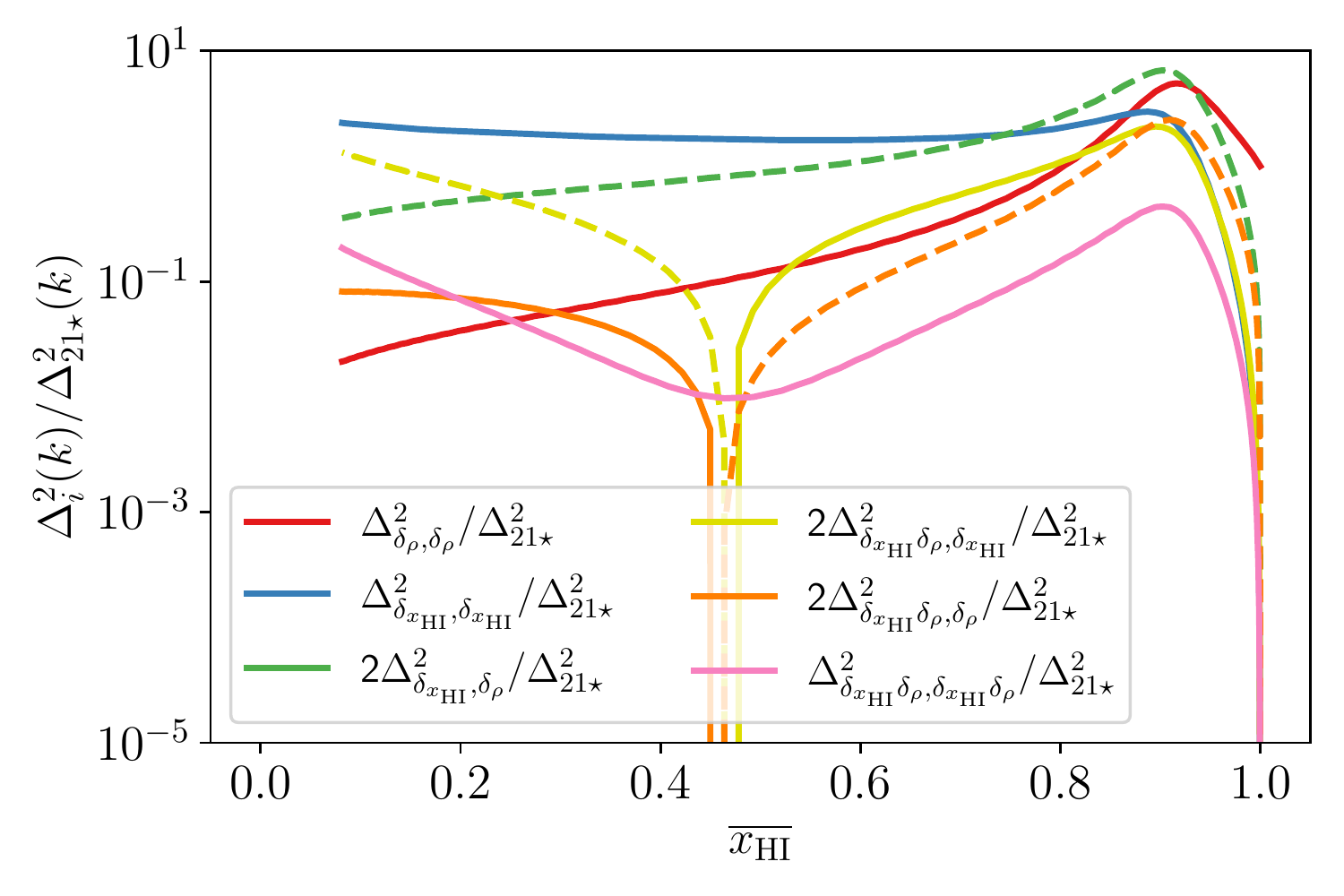}
\caption{Fractional contribution of each constituent term in Eq. \ref{Eq::Decomposition} at $k = 0.1~\mathrm{Mpc}^{-1}$ throughout the EoR. Solid lines mark positive terms, dashed lines represent the absolute value of the negative terms. Upper Panel: The decomposition of the fiducial 21FR20 model. Lower Panel: Decomposition of the C2RMAX10 model. Note the similarity between the decompositions. We find the shape of this decomposition to be similar for all models explored in this paper.}
\label{fig:21cmFast_DecFrac}
\end{figure}

As can be seen in Fig.~\ref{fig:21cmFast_DecFrac} (upper panel), very early during reionization the density term (red curve) is the dominant component. This is of course hardly surprising given that the ionized fraction is only a few per cent.
The large-scale 21-cm power spectrum during this phase therefore is simply a scaled version of the matter power spectrum, which is why we do not include a plot of it in Fig.~\ref{fig:21cmFast_DecTerms} \citep[but see fig.1 in ][for an example]{Lidz2008}.

\subsection{Equilibration 
\texorpdfstring{(~$0.95>\xb>0.88$~)} )} 
\label{subsec:equilibration}
As soon as the level of ionization exceeds a few per cent, however, the shape of the 21-cm power spectrum starts to change. The nature of this change can be seen in the upper left panel of Fig.~\ref{fig:21cmFast_DecTerms}, where we note that the largest length scales have a very low amplitude, in fact lower than any of the components. The shape of the power spectrum is the result of the contribution of each term, which are of similar magnitude and largely cancel each other. As a consequence it does not resemble the shape of any of the components. Crucially, the signal at large-scales is not matched by either the neutral or density power spectra. The upper panel of Fig.~\ref{fig:21cmFast_DecFrac} shows the rise in the contributions of all components and how they all attain a local maximum of similar magnitude. This relatively brief period of the suppression of large-scale power was first noted by \citet{Lidz2008} who named it ``equilibration''.

During this period, the introduction of new ionizing sources in the densest regions causes the $\delta_{\HI}$ and $\delta_{\rho}$ to gradually become more anti-correlated on the largest scales. The ionization within the densest regions diminishes the contribution of the highest $\delta_{\rho}$ peaks to the 21-cm fluctuations, making them similar to those of the average universe. The cross-power spectrum expresses this cross-correlation and captures this phenomenon. In Fig.~\ref{fig:21cmFast_DecFrac} (upper panel), we observe that the cross term (dashed green line) grows in amplitude and becomes a dominant component. $\Delta_{\delta_{\mathrm{\rho}},\delta_{\HI}}^{2}$ can be viewed as a measure of how well the ionized regions follow the density structures from which they originate. A stronger anti-correlation corresponds to a larger negative contribution of the cross power in the decomposition, causing the dimming of the 21-cm signal on large-scales (black curve) plotted in the top left panel of Fig.~\ref{fig:21cmFast_DecTerms}.

\subsection{Equilibration to mid-point
\texorpdfstring{(~$0.88>\xb>0.50$~)} )} 
\label{subsec:eq2mid}
After the period of equilibration the amplitude of the large-scale $\Delta_{\mathrm{21cm}}^{2}$ is initially dominated by the cross term 
but the neutral term
quickly takes over as the dominant component ($\xb\la 0.80$, see Fig.~\ref{fig:21cmFast_DecFrac} (upper panel)). However, throughout this entire period the shape of the large-scale 21-cm power spectrum is determined by the shape of the neutral component. As can be seen from Fig.~\ref{fig:21cmFast_DecTerms} (upper right panel), this shape develops a non-monotonic character, associated with the presence of ionized regions. The figure also shows that the second most dominant term, the cross term, is a scaled copy of the neutral term. Its relation to the density fluctuations results in a lower amplitude.

\subsection{Mid-Point to End of the EoR
\texorpdfstring{(~$\xb<0.5$~)} )}

During this final period the neutral term
continues to dominate the evolution of the 21-cm power spectrum, there are however significant changes in the contributions of the cross-term and the higher-order terms. At the midpoint of the EoR, $\xb \approx 0.5$ of Fig.~\ref{fig:21cmFast_DecFrac} (upper panel), the higher-order terms (yellow and orange curves) undergo a second process of equilibration. This transition is best understood by comparing the upper right and lower left panels of Fig.~\ref{fig:21cmFast_DecTerms}. 
The $\Delta_{\delta_{\rho}\delta_{x_{\mathrm{HI}}},\delta_{x_{\mathrm{HI}}}}^{2}$ term changes from positive to negative, and $\Delta_{\delta_{\rho}\delta_{x_{\mathrm{HI}}},\delta_{\rho}}^{2}$ changes from negative to positive, both starting at the largest scales. The latter remains negative at small scales though, even at much later times (see the lower right panel). 

The evolution of these terms can be understood by examining the evolution of the neutral perturbation $\delta_{\HI} = \xHI/\xb  -1$. Its minimum value is -1, corresponding to ionized cells. However, its maximum value $\delta_{\HI} = 1/\xb -1$, attained in fully neutral cells, keeps increasing as reionization proceeds. The higher-order terms are particularly sensitive to this change in the range of values for $\delta_{\HI}$.
In terms of the decomposition, both higher-order terms grow in amplitude during this period, with $\Delta_{\delta_{\rho}\delta_{\HI},\delta_{\HI}}^{2}$ evolving faster due to the dominance of the $\delta_{\HI}$ over $\delta_{\rho}$ . Past $\xb \approx 0.4$ (lower right panel of Fig.~\ref{fig:21cmFast_DecTerms}), $\Delta_{\delta_{\rho}\delta_{\HI},\delta_{\HI}}^{2}$ overtakes the cross-term, becoming a second most dominant component of the decomposition. Similarly, the $\Delta_{\delta_{\rho}\delta_{\HI},\delta_{\rho}\delta_{\HI}}^{2}$
gains importance towards the end of reionization (fuchsia line seen in Fig.~\ref{fig:21cmFast_DecFrac} (upper panel)).
\citetalias{Lidz2007} (see eq.~14 and 15 ) derived estimates for
$\Delta^{2}_{\xHI \delta_{\rho},\xHI}$ and $\Delta^{2}_{\xHI \delta_{\rho}, \delta_{\rho} \xHI}$ as functions of  $\Delta^{2}_{\xHI, \xHI}$, valid for large scales in the case of strong $\delta_{\HI}$ fluctuations. We tested these approximations for $k\lesssim 0.3$~Mpc$^{-1}$ and confirmed them to provide a good description of the shape and amplitude of the two higher-order terms for $\xb <0.4$.

The non-monotonicity noted above in Sec. \ref{subsec:eq2mid} has by $\xb =0.233$ developed into a prominent break near $k=0.1~\mathrm{Mpc}^{-1}$ seen in the 21-cm power spectrum and all decomposition components except the density term (Fig.~\ref{fig:21cmFast_DecTerms}, lower right panel). In Sec.~\ref{Sec:MFP_Study} we present a comprehensive study of the parameters impacting the nature of this feature.

\subsection{Generality of the decomposition}
To summarize our findings thus far, up to about 5 per cent ionization the large-scale $\Delta_{\mathrm{21cm}}^{2}$ follows the shape of $\Delta_{\delta_{\rho},\delta_{\rho}}^{2}$. After a transition period known as equilibration and characterized by a suppression of large-scale 21-cm power, its shape is primarily determined by $\Delta_{\delta_{\HI},\delta_{\HI}}^{2}$. During most of the EoR, $\Delta_{\delta_{\rho},\delta_{\HI}}^{2}$ is an important component of the signal, while higher-order terms only become significant towards the later stages. These conclusions were obtained for one particular simulation. In this subsection we explore the generality of these findings.

We start by comparing the 21-cm power spectrum decomposition for {\sc C$^2$-Ray} and {\sc 21cmFAST} simulation results. These two codes employ very different computational approaches (see Sec.~\ref{Sec:Simulations}). Whilst direct comparison is not trivial, a qualitative study of the behaviour of the terms can serve as a consistency check. 
We base our comparison on the evolution of the decomposition terms for our fiducial simulation 21FR20 and C2RMAX10. These two simulations have similar reionization histories (see Figs.~\ref{fig:21cmFAST_Histories} and \ref{fig:C2Ray_History}) and similar source populations with $M_{\mathrm{min}}=10^9 M_{\odot}$. The left panel of Fig.~\ref{fig:21cmFast_DecFrac} (lower panel) shows the evolution of the relative contribution of the decomposition terms at $k=0.1~\mathrm{Mpc}^{-1}$ and can be compared to Fig.~\ref{fig:21cmFast_DecFrac} (upper panel). These two figures are remarkably similar which suggests that the decomposition is not sensitive to the simulation method.

The similarity of the decomposition for these two simulations is in fact not exceptional. We find that the intrinsic shape and evolution of the decomposition is qualitatively similar between all {\sc 21cmFAST} and {\sc C$^2$-Ray} models used in this study. It therefore does not seem to be very sensitive to variations in source efficiency, the minimum mass of sources, or the value of the MFP. Even though the power spectra themselves do differ between the simulations, their decomposition does not vary appreciably.

The only feature which seems to be somewhat sensitive to changes in parameters is the one associated with the short period of equilibration. We find that its magnitude and timing depend on the source population and how strongly the ionizing sources trace the density field. For example, the lower efficiency simulation 21FE10 reaches the maximum of equilibration (lowest power on large-scales) at $\xb = 0.893$, compared to $\xb = 0.917$ for our fiducial model 21FR20 and also attains a lower power at that point. For less efficient sources the ionized bubbles develop at a slower pace, pushing equilibration to lower redshifts. At this point the source population is more numerous which increases the anti-correlation between ionization and the density field. Hence more density peaks are ``hidden'' in the 21-cm signal, suppressing the signal and boosting the decomposition feature.
Similarly, changing the type of source population by modifying the minimum halo mass, also affects the timing of equilibration. For 21FM8, equilibration is reached earlier in terms of redshift but at a lower mean neutral fraction of $\xb = 0.914 $. Whereas 21FM10 reaches it later in time yet at an earlier stage of reionization, $\xb = 0.935 $. The strength of the equilibration feature is not much affected by these differences.

Our results are obviously only valid and general for the assumed physics. For example reionization models in which the sources of ionization are uncorrelated or anti-correlated with the density field, will yield very different decompositions. In fact, some differences with the results outlined above already appear in the case of a weaker correlation caused by Poisson fluctuations in the halo population when only considering rare, very high mass haloes. We present such a case in Appendix~\ref{appendix:C2RBL70}. 

Another case which will change our results is when spin temperature fluctuations contribute substantially to the 21-cm signal. This is more likely to occur at the earlier stages of reionization \citep[e.g][]{Pritchard2007} and thus to affect the equilibration and density dominated stages (subsections \ref{subsec:Early_EoR} and \ref{subsec:equilibration}). Lastly, we did not include redshift space distortions or the light cone effect, which have both been shown to alter the shape of the 21-cm power spectrum \citep[e.g.][]{Barkana2006, Mao2012, Datta2012, Mondal2018,Ross2021}. However, a preliminary check showed no major changes to the evolution of the decomposition when redshift space distortions were included. The light cone effect likely only affects periods of very rapid changes, so possibly the equilibration period. We leave a more in depth study of these effects for future work.

\section{The effect of the mean free path on the large-scale power spectrum}
\label{Sec:MFP_Study}

Our analysis of the decomposition in Sec.~\ref{Sec:DecStudy} indicates that the large-scale $\Delta_{\mathrm{21cm}}^{2}$ after equilibration is a scaled version of $\Delta_{\delta_{\HI},\delta_{\HI}}^{2}$. We have also seen that this neutral term has a non-monotonic shape which can even display a noticeable break (Fig.~\ref{fig:21cmFast_DecTerms} lower right panel). As we will show below, the shape of the neutral term
at large-scales for these {\sc 21cmFAST} simulations is mostly sensitive to the choice of the maximum horizon for ionizing photons parameter $R_{\mathrm{max}}$, which is related to the mean free path of ionising photons (MFP). In this section we will examine the shape of $\Delta_{\delta_{\HI},\delta_{\HI}}^{2}$, the effect of the $R_{\mathrm{max}}$ parameter and its relation with the shape of the density power spectrum and the MFP quantity.

The MFP is the average distance an ionizing UV photon at 13.6 eV will travel through the ionized IGM before it encounters an absorbing structure. Early during reionization this will most likely be the edge of an ionized region but as these regions grow in size chances increase that the absorbing structures will be small scale residual low-density neutral gas located inside the $\ion{H}{II}$ regions \citep[e.g.][]{Bianco2021} or dense clumps with optical depths $>1$, which are often referred to as Lyman limit systems \citep[LLS, see e.g.][]{Alvarez2012, McQuinn2011, 2016MNRAS.458..135S, D'Aloisio2020}. Although the distribution of the large ionized regions is correctly modelled by reionization simulations, the small scale density fluctuations which cause absorption within ionized regions are usually not well resolved. The simulations, therefore, require a subgrid model to limit the distance that ionizing photons can travel inside ionized regions. Below we will refer to the impact of this limited distance that ionizing photons can travel through the ionized IGM as the {\em MFP effect} and the different subgrid models as {\em MFP models} or implementations. The characteristic distance will be referred to as the {\em MFP value} or $\lambda_\mathrm{MFP}$.

Observationally, the value of the MFP is derived from the analysis of the \lya forest spectra towards Quasi-Stellar Objects (QSOs) \citep{Alvarez2007,Worseck2014} near the end of or after reionization, see fig. 9 of \citet{Becker2021} for a overview of different measurements. \citet{Worseck2014} found a value of $\lambda_{\mathrm{MFP}} = 90.64 \pm 14.08$~Mpc at $z = 5.16$ and a scaling of $\lambda_{\mathrm{MFP}} \propto (1+z)^{-6.4}~\mathrm{Mpc}$ with redshift, although the recent work of \citet{Becker2021} claims that this relation does not hold at redshifts relevant for reionization. As the ionized regions can easily reach sizes much larger than this, it is essential for reionization simulations to include the MFP effect as it constitutes a important sink of ionizing photons during the later stages of the EoR.

Implementations of the MFP effect in reionization simulations fall into two categories, namely (hard) barriers which impose a maximum distance out to which ionizing photons from sources can travel, which is then equated to the MFP value, or a more gradual absorption of ionizing photons in which a fraction $e^{-1}$ remains after a distance $\lambda_\mathrm{MFP}$. Appendix \ref{appendix:MFP_application} provides an overview of how different reionization simulations have included the MFP effect. Our {\sc 21cmFAST} simulations employ a barrier which is set by the value of the $R_{\mathrm{max}}$ parameter, which therefore is equated to the value of the MFP, see Sec.~\ref{sec:21cmfast}. For our exploration of the impact of the $R_{\mathrm{max}}$ parameter we use a set of four 21FR simulations with $R_{\mathrm{max}}$ values 10, 20, 40 and 70~Mpc. We base this investigation on the scale-dependent bias $b(k)$, as defined in Eq.~\ref{eq:Bias}. After considering the results for different $R_{\mathrm{max}}$ values, we also study the impact of the other simulation parameters on $b(k)$, as well as how different implementations of the MFP affect our conclusions.

\begin{figure*}
\includegraphics[width=\columnwidth]{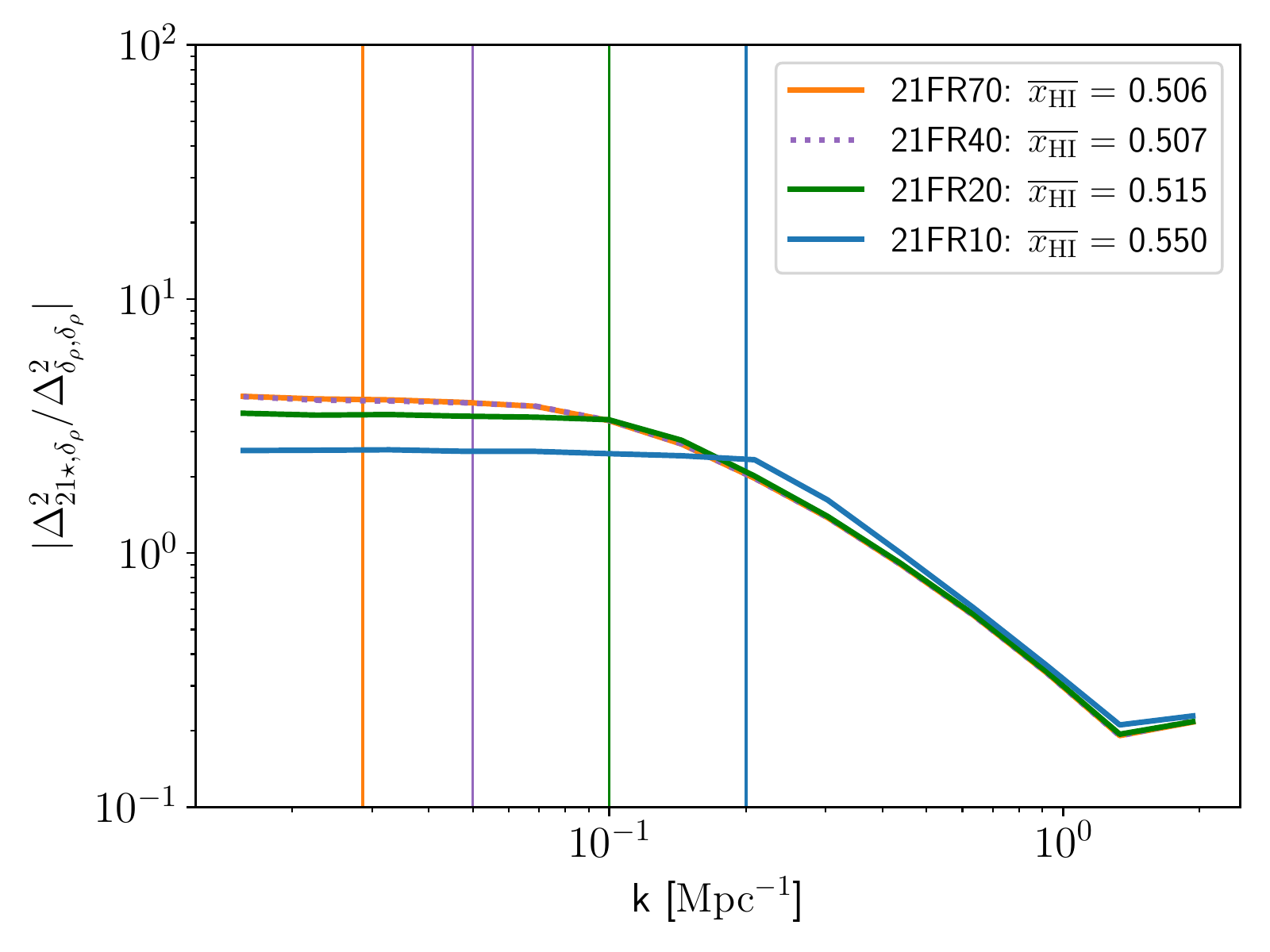}
\includegraphics[width=\columnwidth]{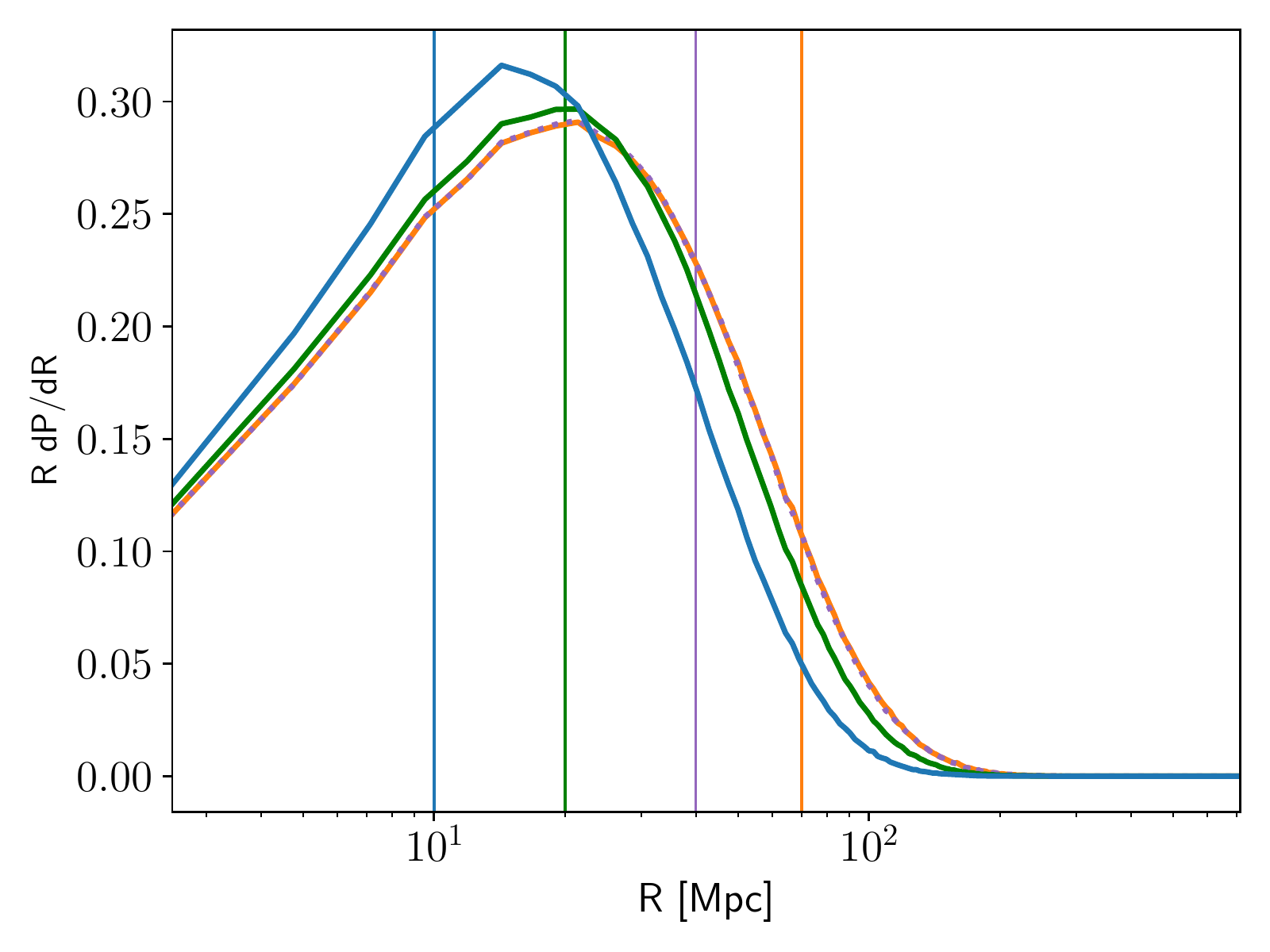}
\caption{Left Panel: Absolute bias values at the midpoint $\xb \approx 0.5$ for the four 21cmFAST simulations with different mean free path values. The corresponding redshifts are $z = 7.6,7.8,7.8,7.8$ for 21FR10, 21FR20, 21FR40 and 21FR70, respectively. Note the sharp transition features at $k_{\mathrm{trans}} \approx 2/R_{\mathrm{max}} = 0.20, 0.10~\mathrm{Mpc}^{-1}$ for 21FR10 and 21FR20, represented by the thin solid vertical lines. The line style of the 21FR40 bias is dotted as it overlaps with that of 21FR70. Right Panel: The corresponding bubble size distributions. The $R_{\mathrm{max}}$ constraint has restricted the growth of ionized regions for 21FR10 and 21FR20 but does not introduce any feature around the $R_{\mathrm{max}}$ values.
}
\label{fig:21cmFASTBias21_xHI_0.5}
\end{figure*}

\begin{figure*}
\includegraphics[width=\columnwidth]{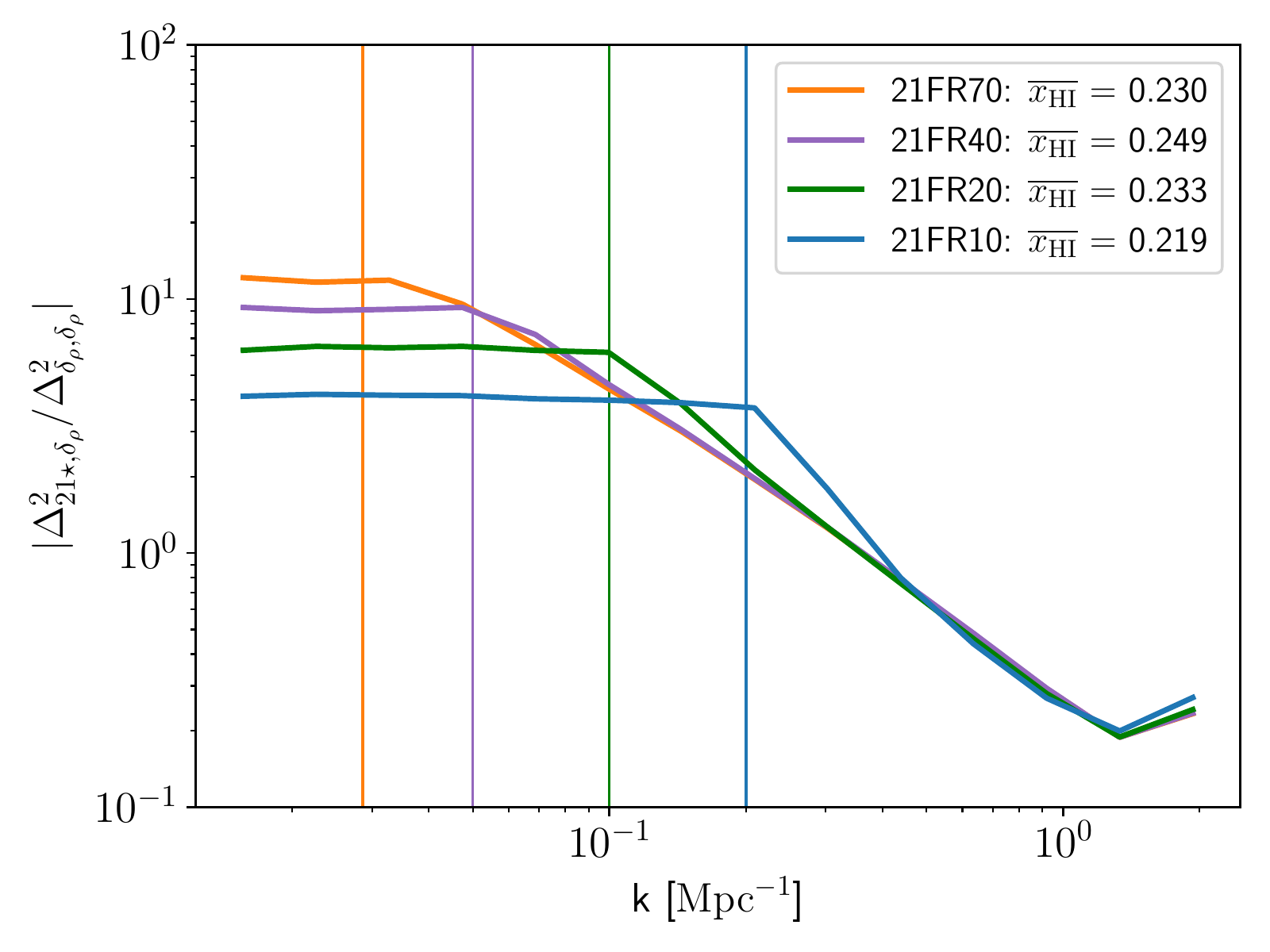}
\includegraphics[width=\columnwidth]{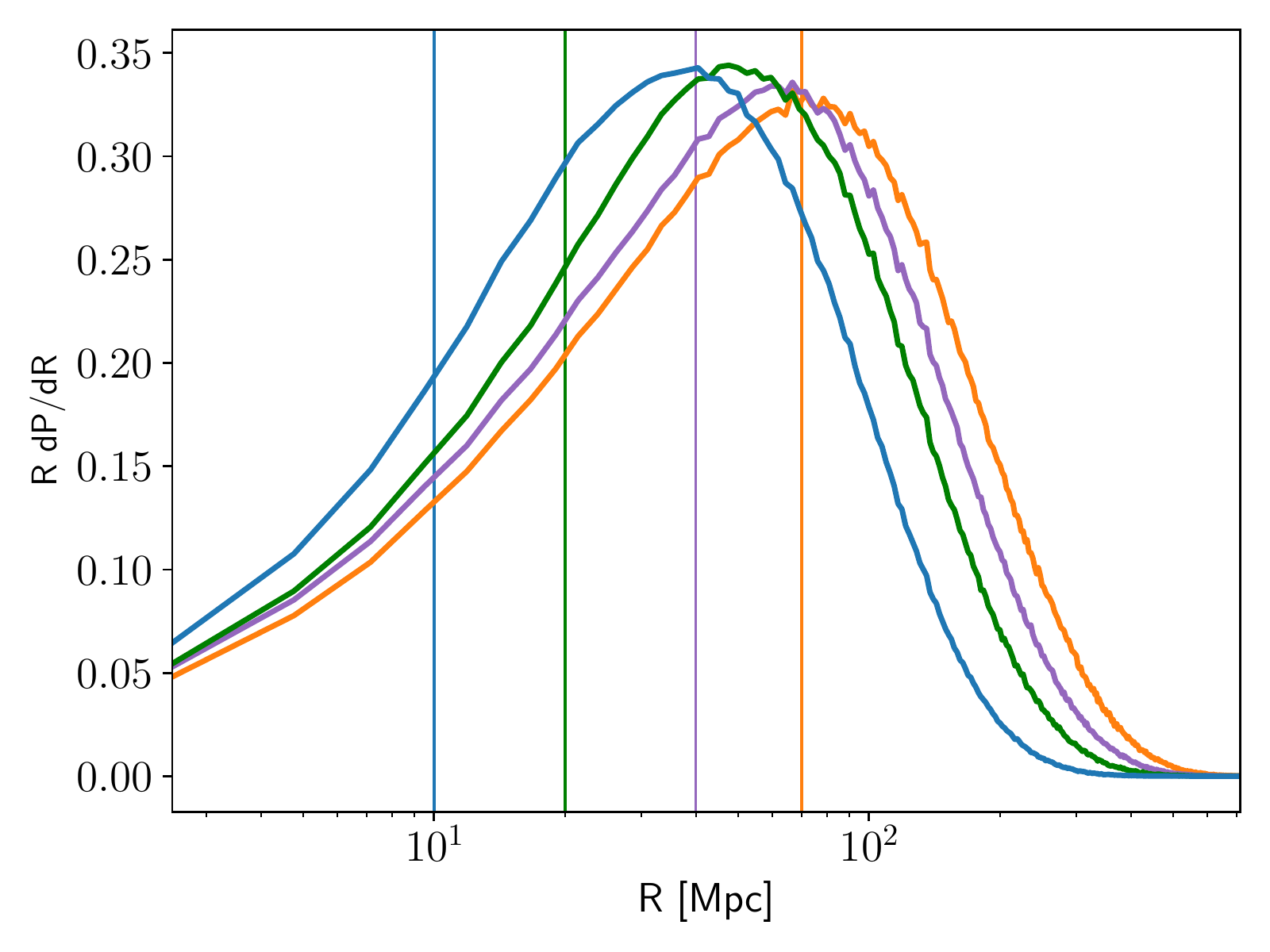}
\caption{As Fig. \ref{fig:21cmFASTBias21_xHI_0.5} but for $\xb \approx 0.2$. The corresponding redshifts are $z = 6.6,7.0,7.2,7.2$ for 21FR10, 21FR20, 21FR40 and 21FR70, respectively. For each of these respective models we notice a transition feature to a scale-independent bias at $k_{\mathrm{trans}} \approx 2/R_{\mathrm{max}} = 0.20, 0.10, 0.05, 0.029~\mathrm{Mpc}^{-1}$ in the left panel, represented by the thin solid vertical lines. At this stage BSDs are different for all four models.
}
\label{fig:21cmFASTBias21_xHI_0.2}
\end{figure*}

\subsection{\texorpdfstring{\textbf{$R_{\mathrm{max}}$}}p and the large-scale bias}
\label{sec:mfp_and_bias}
As we saw above in Sec.~\ref{subsec:Early_EoR}, the large-scale 21-cm signal traces the matter density field prior to equilibration. As a consequence, the large-scale bias $b(k)$ during the earliest phases is positive and scale-independent. After the equilibration period, sources have completed ionizing their locally dense environments. As the neutral hydrogen field is now anti-correlated with the density field, the $\Delta_{21 \star,\delta_{\rho}}^{2}$ term and hence $b(k)$ flips from positive to negative. We find that the large-scale bias remains negative until the end of reionization. Hence, it is represented by its absolute in all figures in this section. 

We now consider the bias parameter in Fig.~\ref{fig:21cmFASTBias21_xHI_0.5} (left panel) at the midpoint of reionization
for all four 21FR models in which the $R_{\mathrm{max}}$ is set to 10, 20, 40 and 70~Mpc. Starting at high $k$ (small scales), we see that the bias has a scale-dependent regime down to a certain $k$-value, beyond which it becomes scale-independent. Both the bias parameters of 21FR40 and 21FR70 are identical and overlap (hence the 21FR40 bias is shown with a dotted line style). Their transition point occurs at $k\approx 0.08~\mathrm{Mpc}^{-1}$. However, for the simulations with $R_{\mathrm{max}} = 10$ and 20~Mpc, the transition occurs at the higher $k$-values of $k = 0.2$ and $0.1~\mathrm{Mpc}^{-1}$, respectively. As we saw in Sec.~\ref{Sec:DecStudy}, at this stage the large-scale 21-cm signal follows the shape of neutral fraction power spectra. However the bias results show that this neutral term, and hence the 21-cm power spectrum, beyond a certain length scale actually follows the density term. These results show that the transition scale depends on the MFP for ionizing photons, which for 21FR10 and 21FR20 is already set by the subgrid MFP model (here the value of $R_\mathrm{max}$) but for 21FR40 and 21FR70 is still set by the sizes of ionized regions. For the former models the relation between $R_\mathrm{max}$ and the transition scale $k_\mathrm{trans}$ is found to be $k_{\mathrm{trans}} \approx 2/R_{\mathrm{max}}$. The vertical lines in Fig.~\ref{fig:21cmFASTBias21_xHI_0.5} indicate this scale for the different simulations. 

The right hand panel of Fig.~\ref{fig:21cmFASTBias21_xHI_0.5} shows the corresponding bubble size distributions (hereafter referred to as BSD), where the distribution's peak characterizes the typical size of ionizing regions \citep[e.g.][]{giri2018a}. These BSDs have been calculated by the method proposed in \citet{Mesinger2007}.\footnote{This method is commonly called the MFP method, although it should be pointed out that it is not related the MFP parameter which we are considering in this paper.} While the BSDs for the simulations with $R_{\mathrm{max}} \ge 40$~Mpc are nearly identical, the two simulations with shorter $R_{\mathrm{max}}$ values peak at smaller bubble sizes. Note that the $R_{\mathrm{max}}$ value (indicated with vertical lines for the different simulations) for 21FR10 ($R_{\mathrm{max}} = 10 $~Mpc) does not inhibit the formation of ionized regions larger than that and the peak of the BSD lies in fact closer to 20 Mpc. Overlap of adjacent ionized regions obviously allows the formation of ionized regions larger than the $R_{\mathrm{max}}$ value and as a consequence the BSD does not show any clear feature related to it
However, as we saw above, the bias parameter clearly does. For the remaining 21FR simulations, the $R_{\mathrm{max}}$ constraints are much larger than the characteristic bubble size. As regions larger than $R_{\mathrm{max}}$ have not had time to form, there is no significant suppression of them 
and the BSDs are identical.

The larger ionized bubbles, seen at the high-end tail of the BSD of the right panel of Fig.~\ref{fig:21cmFASTBias21_xHI_0.5}, tend to be the result of overlapping bubbles at the densest regions. Hence, scales larger than the $R_{\mathrm{max}}$ in this period are occupied by large \Hi~regions, which follow the large-scale density distribution. Therefore the large-scale structure of the 21-cm signal is defined by the distribution of neutral regions, while its small-scale is governed by that of the ionized bubbles. We observe this effect in the shape of the bias parameter from Fig.~\ref{fig:21cmFASTBias21_xHI_0.5} (left panel).
The large-scale biases of the 21FR10 and 21FR20 models are scale-independent up to the transition feature, beyond which the bias becomes scale-dependent. In fact, all simulations have such a scale-independent regime, the transition point of which appears to depend on the true MFP, either set by the typical size of the ionized regions or by the $R_{\mathrm{max}}$ value, depending on which of the two is smaller. 

\begin{figure}
	\includegraphics[width=\columnwidth]{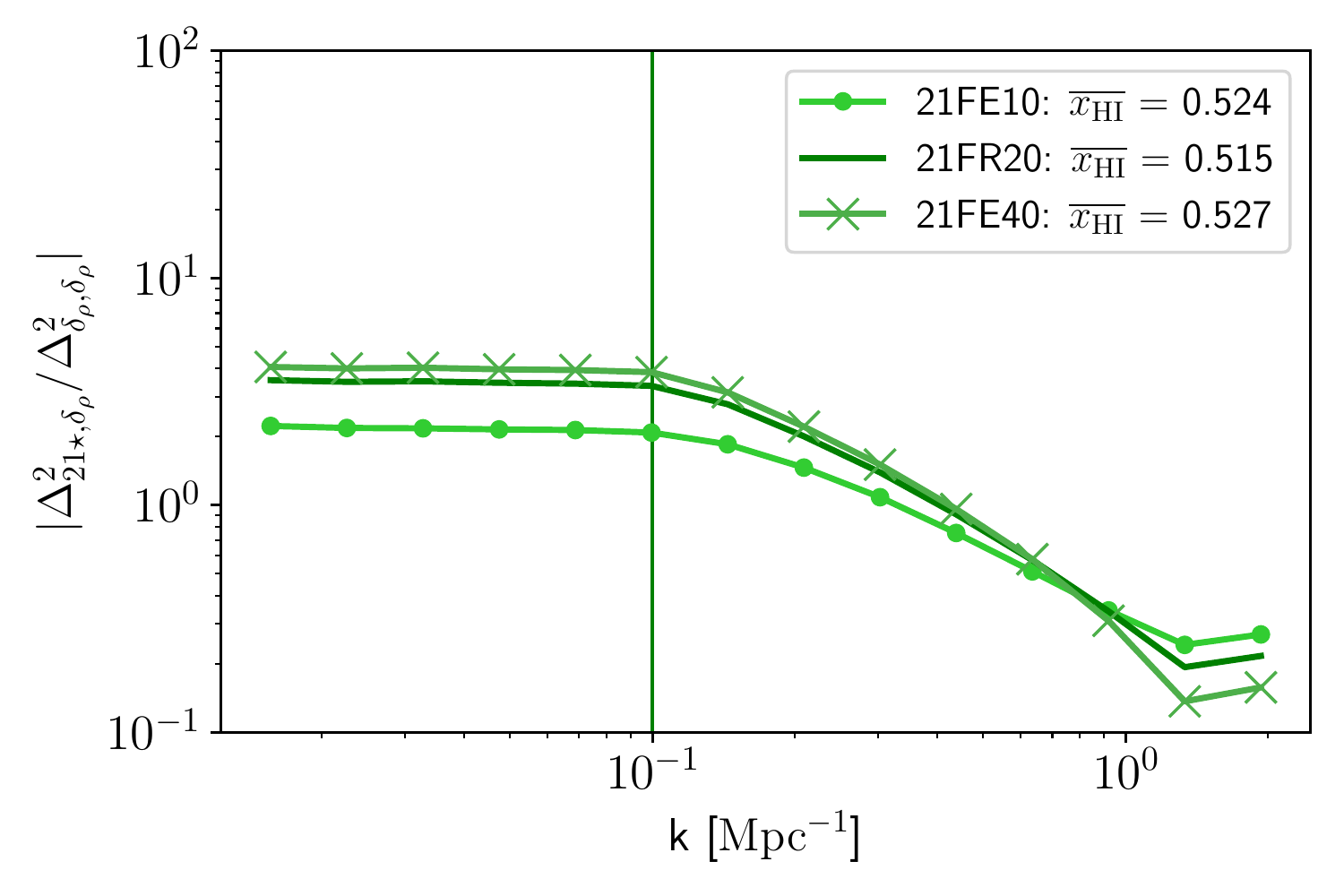}
	\includegraphics[width=\columnwidth]{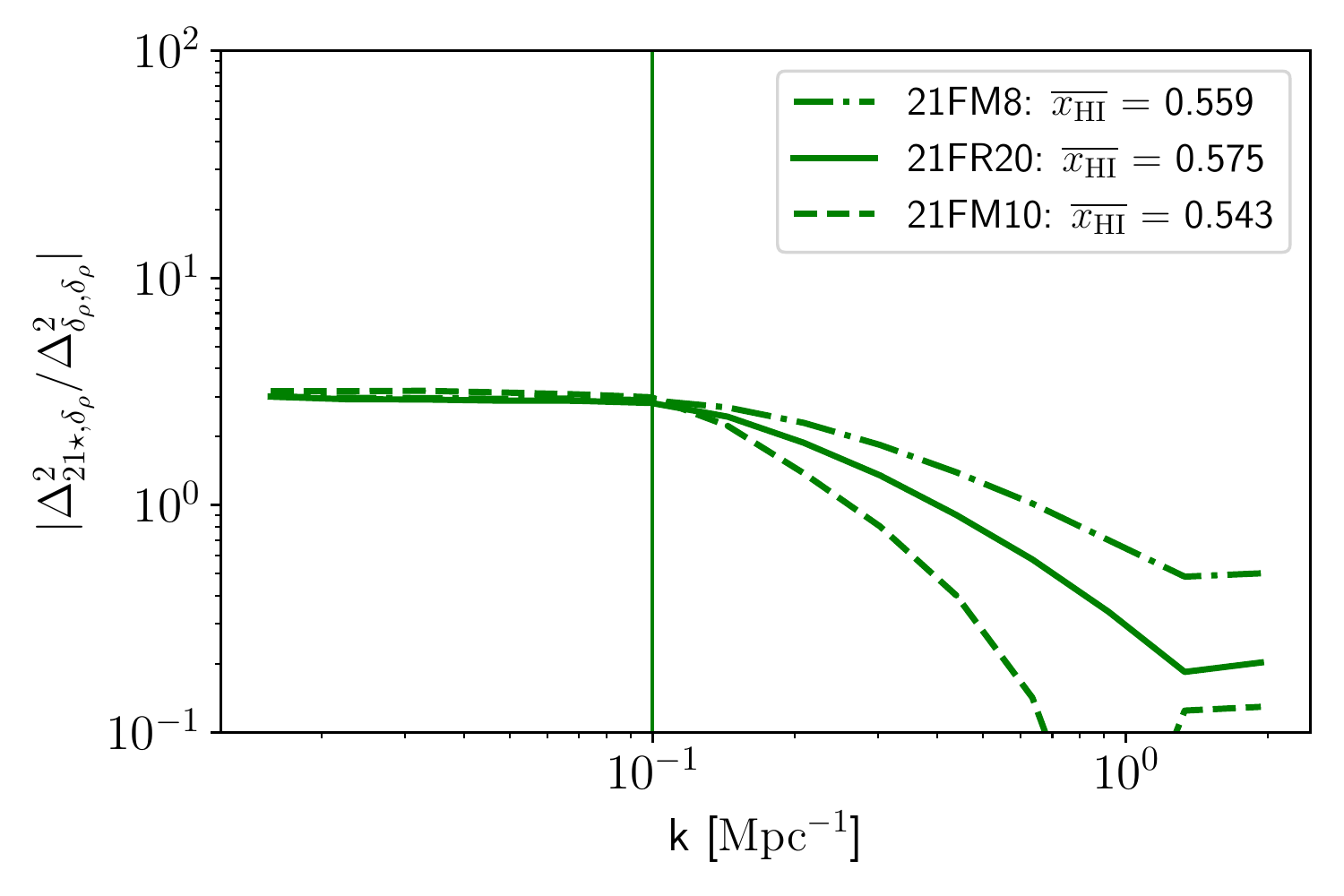}
    \caption{Absolute bias values at the midpoint of reionization for a set of {\sc 21cmFAST} simulations with identical mean free path values, where 21FR20 is chosen as the fiducial model. The model parameters are described in more detail in Tab.~\ref{tab:21cmSimulations}.
    Upper Panel: Bias comparison of three models with a varying efficiency parameter $\zeta$. The redshift values are $z = 6.2,7.8,8.6$ for 21FE10, 21FR20 and 21FE40, respectively. Lower Panel: Bias comparison of three models where the minimal mass $M_{\mathrm{min}}$ is varied. The redshift values are $z = 10.0,8.0,6.0$ for 21FM8, 21FR20 and 21FM10, respectively. In essence we observe the transition feature for all models used in this figure to be at $k_{\mathrm{trans}} \approx 2/R_{\mathrm{max}} = 0.10~\mathrm{Mpc}^{-1}$, visualised by the thin solid vertical lines . For the scale-independent regime of the bias parameter, altering the ionization efficiency shifts the amplitude. Conversely, changes in minimal mass does not affect the large-scale shape or amplitude.
    }
    \label{fig:21cmFast_EM_Bias_Evar}
\end{figure}

The transition feature becomes more pronounced at the later stages of the EoR. Fig.~\ref{fig:21cmFASTBias21_xHI_0.2} (left panel) highlights this phenomenon, where
the bias values are compared at a neutral fraction of $\xb \approx 0.2$. All models now clearly exhibit different transition scales but follow similar trends in the scale-dependent regime above $ k \approx 0.3~\mathrm{Mpc}^{-1}$. The measured transition scales are 0.20, 0.10, 0.05 and 0.033 Mpc$^{-1}$ for 21FR10, 21FR20, 21FR40 and 21FR70, respectively. This shows that this scale has not changed since $\xb \approx 0.2$ for simulations 21FR10 and 21FR20 and for 21FR40  it confirms the $k_{\mathrm{trans}} \approx 2/R_{\mathrm{max}}$ relation. For the 21FR70 simulation the true MFP is still limited by the sizes of ionized regions. Results for later stages indeed confirm that its value of $k_{\mathrm{trans}}$ further decreases until it reaches $k_{\mathrm{trans}} = 2/70 \approx 0.029$~Mpc$^{-1}$.

The BSDs for this later period are presented in Fig.~\ref{fig:21cmFASTBias21_xHI_0.2} (right panel) and are seen to diverge from each other. Comparing to Fig.~\ref{fig:21cmFASTBias21_xHI_0.5} (right panel), we note the BSDs of 21FR10, 21FR20 are slower to evolve, as the growth of their ionized regions is already affected by their smaller $R_{\mathrm{max}}$ values by the time the midpoint of the EoR is reached. The $R_\mathrm{max}$ parameter restricts the growth of all $\HII$ regions of comparable size throughout the EoR. However, when the peak of the BSD reaches sizes comparable to $R_\mathrm{max}$, the growth of the majority of the ionized bubbles in the distribution becomes suppressed, and the transition feature becomes more pronounced. Analysis of both earlier and later stages (not shown) reveal that for 21FR40 the peak of the BSDs has only recently grown to scales comparable to its $R_{\mathrm{max}}$ value, and now starts to deviate from the 21FR70 BSD. 

In conclusion, we find that at any given stage of reionization, the bias parameter has a scale-independent (linear) regime for lower $k$ values and a scale-dependent regime for higher $k$ values. The scale at which one regime
transitions to the other, $k_\mathrm{trans}$, is found to depend on the actual MFP of ionizing photons, which initially is set by the typical size of ionized regions. Once the bubbles grow larger than the limit imposed by the implementation of the MFP effect (in this case $R_\mathrm{max}$) , $k_\mathrm{trans}$ becomes equal to 2/$R_\mathrm{max}$ and stops evolving. 
For the smaller $R_{\mathrm{max}}$ models, such as 21FR10 and 21FR20 this has already happened by the midpoint of reionization.

We explored $R_{\mathrm{max}}$ values ranging from 10 to 70~Mpc. However, recent results of \citet{Becker2021}, claim a value of value of $\lambda_{\mathrm{MFP}} = 5.25 \pm 4.55 $~Mpc at z = 6, hinting at a short MFP and a more rapid evolution at the late stages of the EoR. Work by \citet{Cain2021} and \citet{Davies2021b} reconcile these measurements with rapid models of reonization with enhanced sinks, which extend reionization to z = 5 and show the considerable impact a low value of the MFP has on the reionization process. 

\begin{figure*}
	\includegraphics[width=1.9\columnwidth]{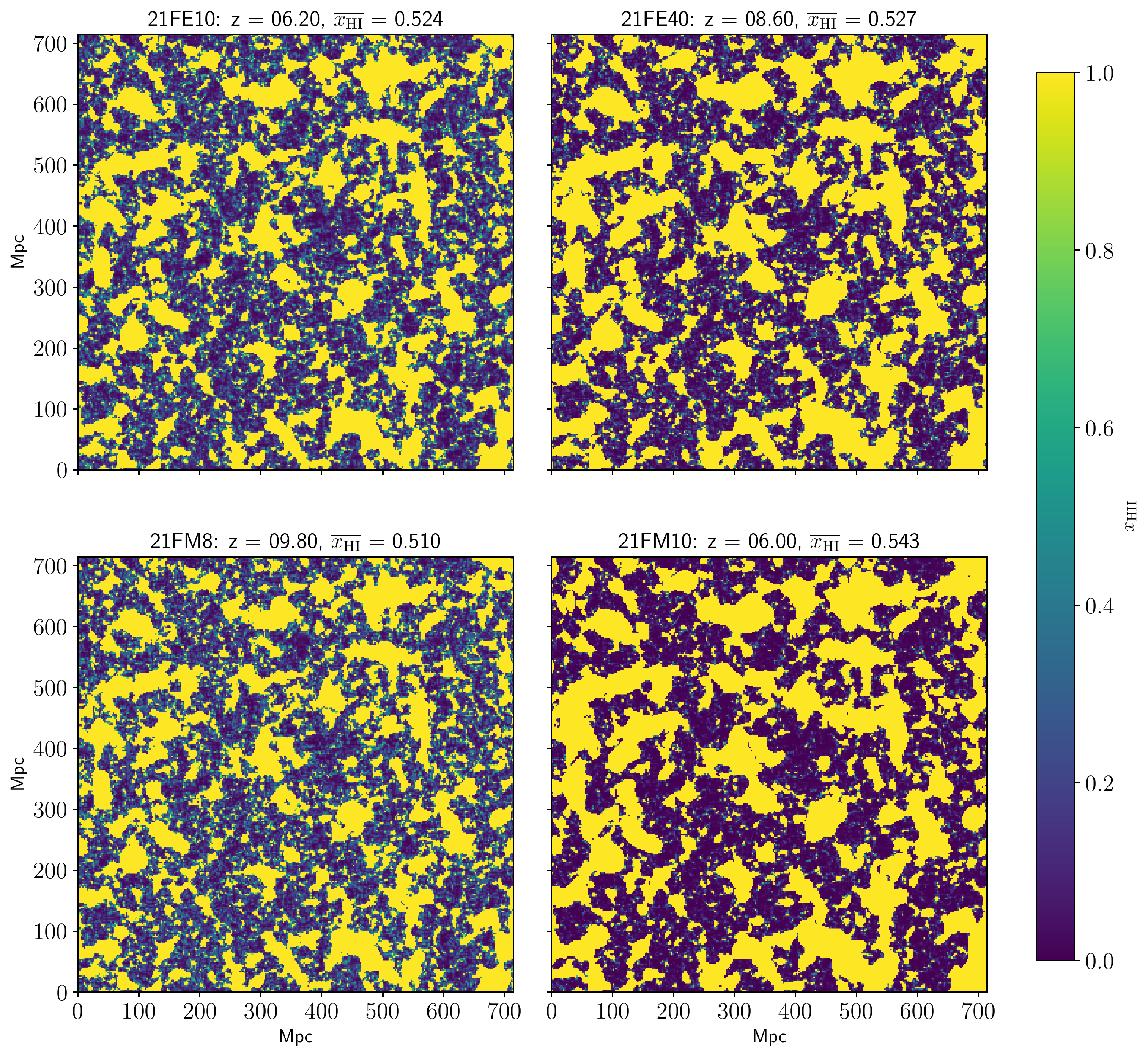}
    \caption{Ionization fraction slices at the midpoint ($\xb{} \approx 0.5$) of the EoR for a selection of {\sc 21cmFAST} simulations. Upper panels: We examine the effect of varying the efficiency parameter $\zeta$ for the 21FE10 (left panel) and 21FE40 (right panel). Lower panels: We compare the effect of the minimal mass $\mathrm{M_{\mathrm{min}}}$ between 21FM8 (left panel) and 21FM10 (right panel). The simulation parameters are given in Tab.~\ref{tab:21cmSimulations}.}
    \label{fig:21cmFast_EM_Slice}
\end{figure*}

\subsection{Influence of \texorpdfstring{$M_{\mathrm{min}}$}p and \texorpdfstring{$\zeta$ }p on the large-scale bias}

After having studied the effect of the $\mathrm{R_{\mathrm{max}}}$ parameter, we now want to consider the impact of the other two astrophysical parameters on the bias. For this purpose we adopt 21FR20 as our fiducial model, fix the value of
$R_{\mathrm{max}}$ to 20~Mpc, and vary the efficiency and the minimal mass of the haloes contributing to reionization. The set of parameters chosen for this comparison are displayed in Tab.~\ref{tab:21cmSimulations}. 

In Fig.~\ref{fig:21cmFast_EM_Bias_Evar} (upper panel) we plot the bias parameter for three models with varying efficiency $\zeta$: 21FE10, 21FR20 (which for this comparison could be labelled as 21FE25) and 21FE40 at the midpoint of the EoR. In this version of {\sc 21cmFAST}, the $\zeta$ parameter has no mass dependence and so for lower efficiency models at a given redshift the same sources will have produced smaller ionized regions. This implies that models with lower efficiency reach a given stage of reionization at a lower redshift. The relevant redshift values are listed in the caption of Fig.~\ref{fig:21cmFast_EM_Bias_Evar} as well as in the labels of the upper panels of Fig.~\ref{fig:21cmFast_EM_Slice}, which show slices from the midpoint of reionization for each of the three simulations. At lower redshifts, more haloes will have formed, also in less dense regions.
While both large and small scale structures in Fig.~\ref{fig:21cmFast_EM_Slice} look similar between 21FE10 and 21FE40, it can be noted that the large ionized features in 21FE40 (upper right panel) are on average slightly more developed than in 21FE10. On the other hand, more ionization is present in the low density regions of 21FE10. Hence, the overall shape of the bias, seen in the upper panel of Fig.~\ref{fig:21cmFast_EM_Bias_Evar}, is similar on all scales for all models and all show the same transition scale of about 0.1~{Mpc}$^{-1}$. However, the amplitude, especially on the large-scales, is higher for the simulations with larger efficiencies.

Fixing the efficiency and instead changing the minimal halo mass $M_{\mathrm{min}}$ affects the bias in a very different way, as can be seen in Fig.~\ref{fig:21cmFast_EM_Bias_Evar} (lower panel).
Here we compare the 21FM8, 21FR20 (which could also be called 21FM9) and 21FM10 models which only differ in the value for the minimal mass of haloes contributing to reionization. As the simulation names suggest these are $\mathrm{M}_{\mathrm{min}} = 10^{8}, 10^{9},~\mathrm{and}~ 10^{10}~\mathrm{M}_{\odot}$, respectively. For a higher minimum mass, fewer haloes are active at a given redshift and yielding a less reionized universe. This implies also that for a higher minimum mass a given stage of reionization is reached at a lower redshift, as can be seen in the labels of the lower two panels of Fig.~\ref{fig:21cmFast_EM_Slice}.
The bias results at the midpoint of reionization shown in the lower panel of Fig.~\ref{fig:21cmFast_EM_Bias_Evar} show that the position of the transition scale remains the same in all simulations, and is thus solely determined by the $\mathrm{R_{\mathrm{max}}}$ value. However, in contrast to what we saw above, the 21FM simulations produce an identical large-scale bias value down to the transition scale but for 
higher values of the minimal mass yield lower bias values at the small scales. Comparing the ionization slices for 21FM8 and 21FM10, shown in the lower panels of Fig.~\ref{fig:21cmFast_EM_Slice}, we note similar large-scale features between the two models. However, 21FM10 (lower right panel) clearly has fewer small ionized regions in between the larger ones. This is because the lower mass haloes which are producing these features in 21FM8 are not active in 21FM10. This results in a decrease of the small-scale bias amplitude for 21FM10, shown by the dashed line in the lower panel of Fig.~\ref{fig:21cmFast_EM_Bias_Evar}, compared to the value of 21FM8 (dash-dotted).

Note that we find that the large-scale 21-cm bias values of the 21FM8, 21FR20, and 21FM10 models are all similar. Previous work found that the large-scale bias should be mainly determined by the bias of the ionizing sources  \citep[e.g.][]{McQuinn2018} which at a given redshift is of course quite different for these different scenarios. However, when we checked the halo bias values using the analytical model \citep{Tinker2010} from the {\sc Colossus} toolkit \citep{Diemer2018} for these models, we found
that these three halo populations in fact have the same bias at the redshift when each model reaches the midpoint of the EoR. We thus also confirm that our large scale 21-cm bias values depend on the halo bias of the ionizing sources. We intend to investigate this relation further in future work. 

In conclusion, the EoR models with a higher ionization efficiency produce larger bubbles and boost the amplitude of the large-scale bias at the middle-stages of the reionization history. Conversely, varying the minimal mass does not affect the large-scale bias, but changes ionization on small-scales.

\subsection{Impact of the mean free path implementation}
\label{sec:mfp_impact}
\label{subsec:C2_Model_Disuscussion}
As we have seen above, for the {\sc 21cmFAST} simulations the ${R_{\mathrm{max}}}$ parameter leaves a clear signature in both the power spectra (see e.g. the lower right panel of Fig.~\ref{fig:21cmFast_DecTerms}) and the bias (see e.g. the left panel of Fig.~\ref{fig:21cmFASTBias21_xHI_0.2}). Such a sharp transition in these Fourier domain quantities can be expected as in this version of {\sc 21cmFAST}, the ${R_{\mathrm{max}}}$ is implemented as a single sharp cut-off in the Fourier domain. Recent work by \citet{Davies2021} also showed how the use of the $R_{\mathrm{max}}$ parameter in semi-numerical codes which use the excursion set formalism, affects the large-scale power spectrum. They propose two alternative methods which implement the MFP effect as a gradual absorption which result in smoother power spectra.

We investigate the effect of the MFP implementation on the shape of the power spectra and bias by using a set of 
{\sc C$^2$-Ray} simulations. Due to its nature as a ray-tracing code, {\sc C$^2$-Ray} has to implement the MFP effect in real space rather than Fourier space. We provide a brief overview of different methods to implement the MFP effect in Appendix \ref{appendix:MFP_application}.

\begin{figure*}
\includegraphics[width=.49\linewidth]{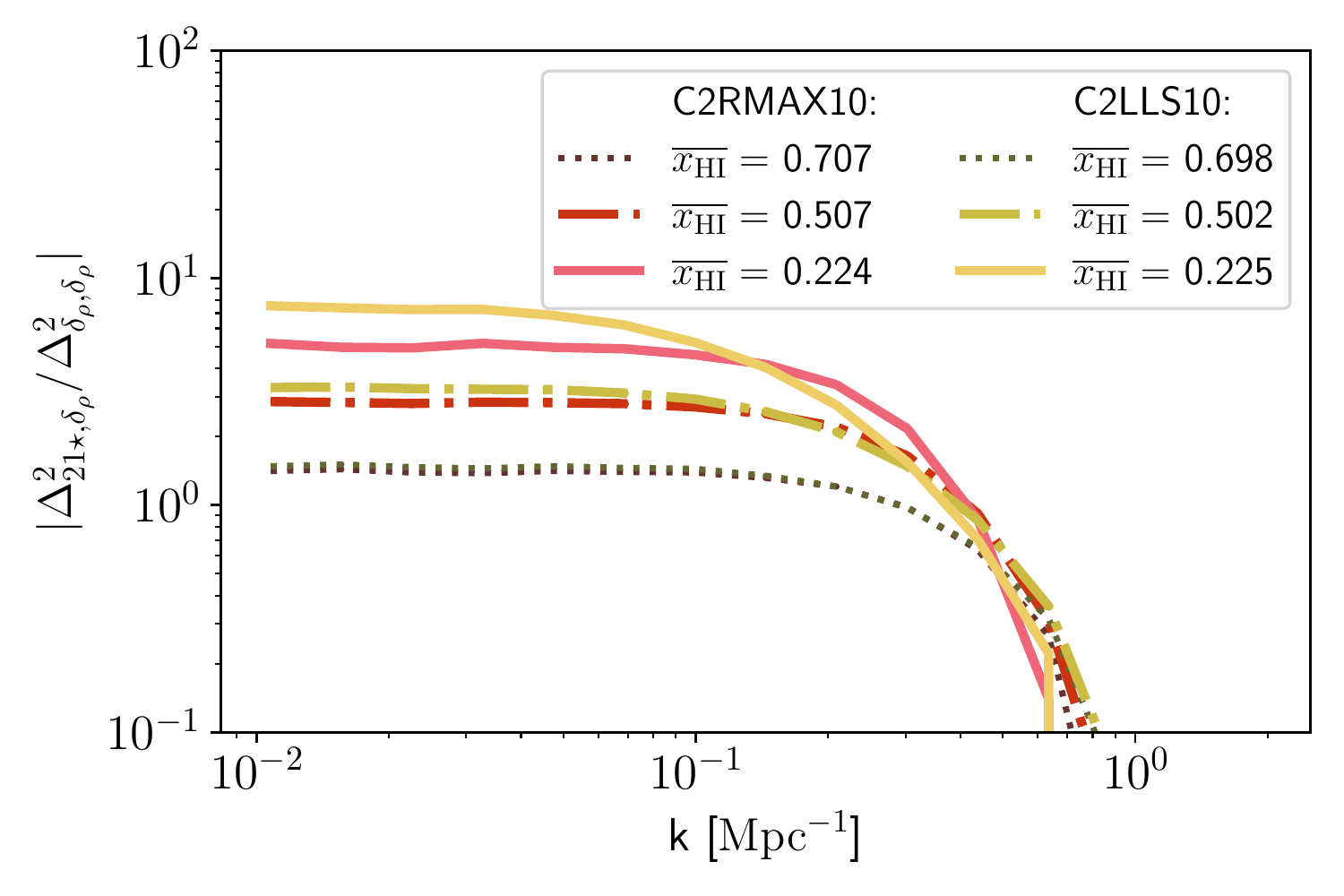}
\includegraphics[width=.49\linewidth]{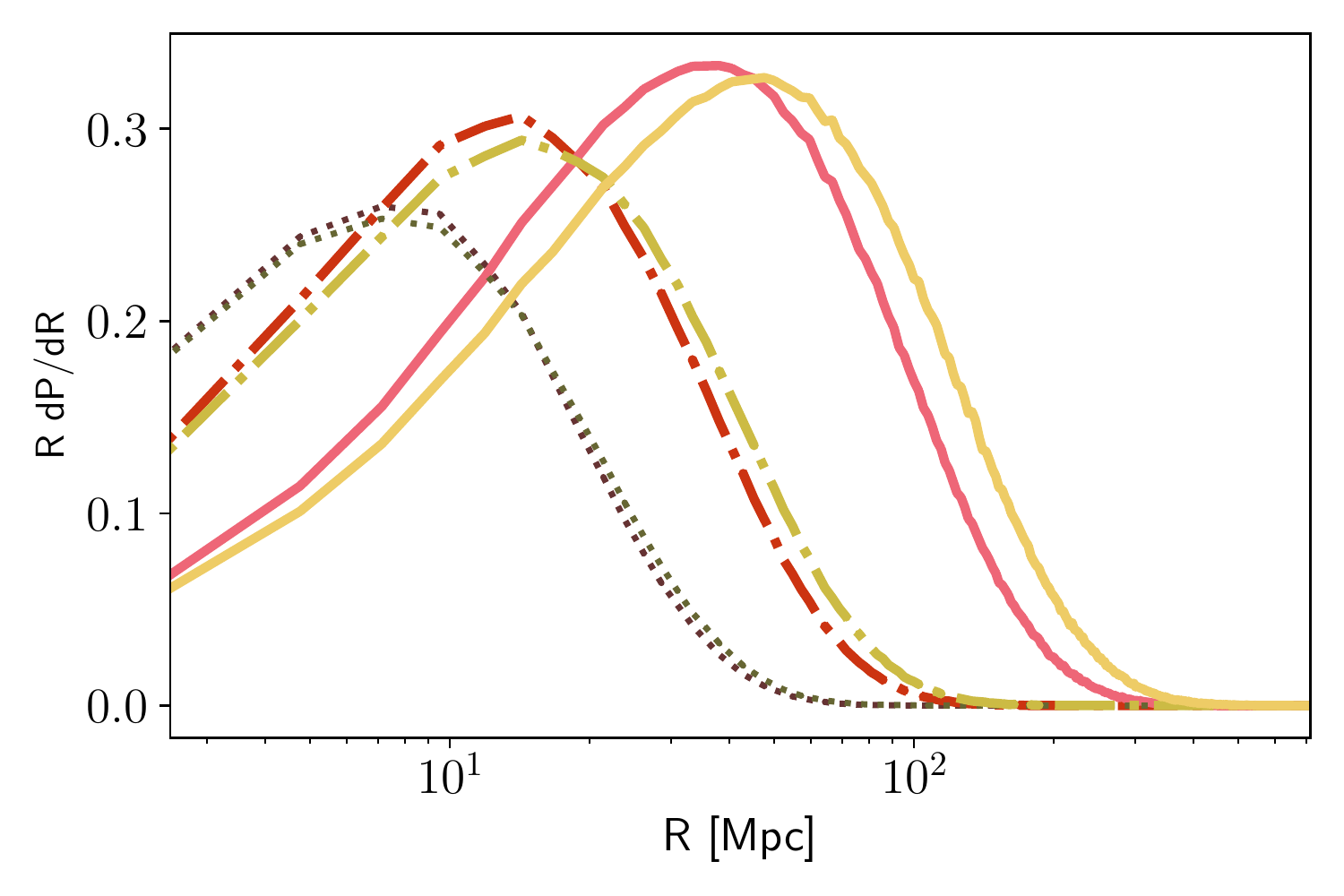}

\caption{Left panel: Absolute value of the bias for simulations C2RMAX10 (red) and C2RLLS10 (yellow) for three distinct stages of reionization, $\xb{} \approx 0.7$ (dotted lines), 0.5 (dash-dotted lines) and 0.2 (solid lines). The corresponding redshift values are: early on $z= 8.397,8.283$, midpoint $z = 7.664,7.617$, and late on $z = 6.721,6.905$. While the transition to a scale-independent bias for C2RMAX10 also follows $k_{\mathrm{trans}}\approx 2/\lambda_{\mathrm{MFP}} =2/10 = 0.2 \mathrm{Mpc}^{-1}$, it is smoother than the feature observed in Fig.~\ref{fig:21cmFASTBias21_xHI_0.2}, due to the implementation of the photon barrier in real space. Conversely the use of an absorption method for the MFP effect in C2LLS10 has smeared out the transition to both lower and higher $k$-scales. Right panel: Corresponding BSDs colour-coded to the appropriate period and simulation.}
\label{fig:C2Ray_21star_Bias}
\end{figure*}

\begin{figure*}
	\includegraphics[width=1.9\columnwidth]{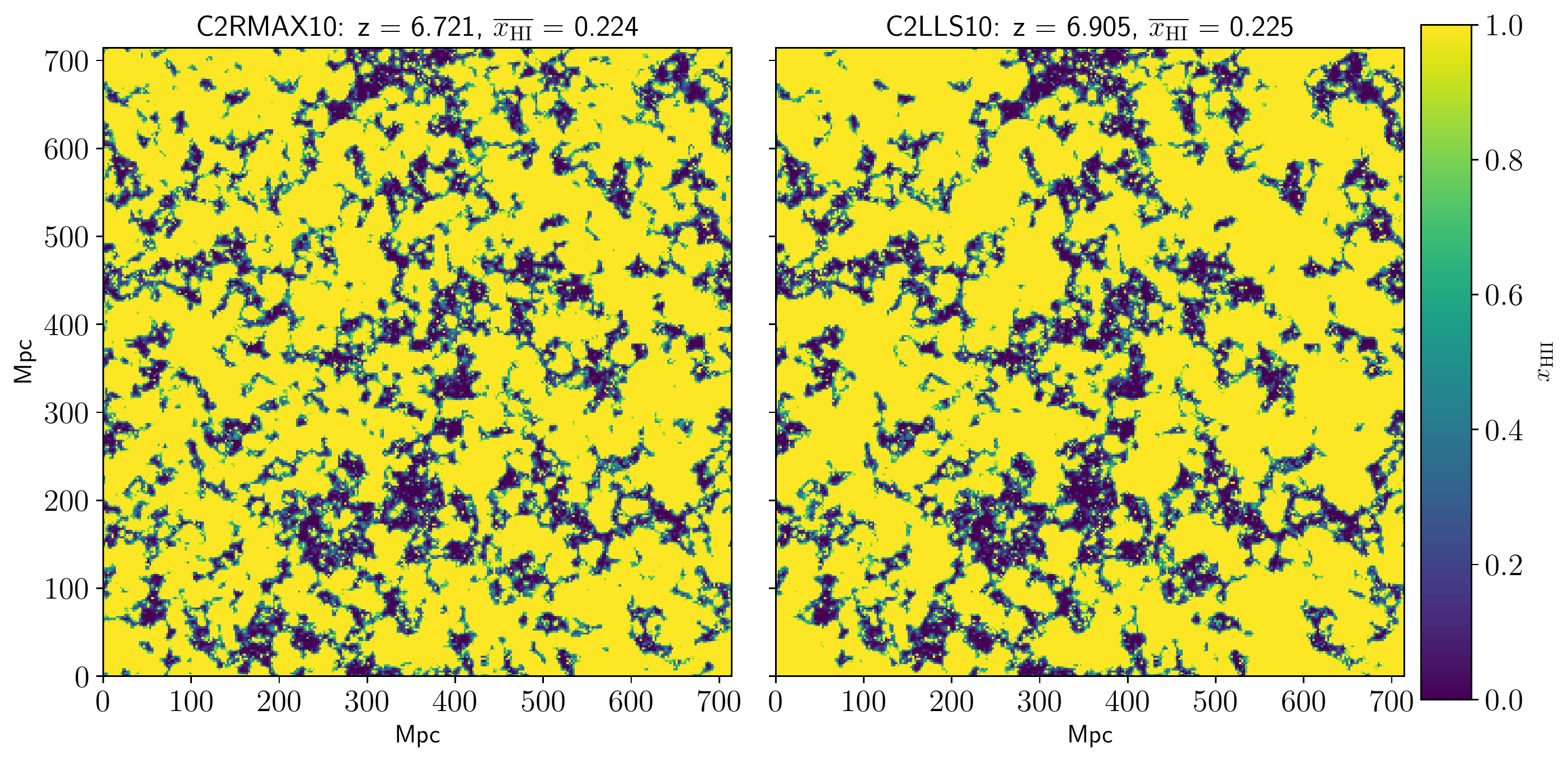}
    \caption{Ionization fraction slices towards the end of the EoR ($\xb{} \approx 0.2$) for the two {\sc C$^2$-Ray} simulations. We compare the effects of using two different MFP implementations, a spherical photon barrier (C2RMAX10, left panel) and continuous absorption (C2LLS10, right panel). Note that for the CLLS10 simulation, ionized bubbles in the densest regions are larger in size, altering the shape of the large-scale bias shown in Fig.~\ref{fig:C2Ray_21star_Bias}. Details of the source parameters are given in Tab.~\ref{tab:C2RaySimulations}.}
    \label{fig:C2Ray_Slice}
\end{figure*}

We start by comparing the bias for two {\sc C$^2$-Ray} simulations, C2RMAX10 and C2LLS10, which only differ in the implementation of the MFP effect. The C2RMAX10 simulation applies a method similar to the $R_\mathrm{max}$ used by {\sc 21cmFAST}. It implements a hard limit for ionizing photons by imposing a spherical barrier of 10~Mpc around each source, beyond which photons are not allowed to propagate. Conversely, the C2LLS10 model adds a small amount of opacity (absorption) to every cell. In the absence of any absorption caused by the actual contents of the cells, an optical depth of $\tau \approx 1$ is reached after a distance of 10~Mpc (see Appendix \ref{appendix:MFP_application}, for more information). This approach is similar to what \citet{Davies2021} developed for {\sc 21cmFAST}.

In Section \ref{subsec:C2_Model_Disuscussion} we showed that the evolution of the decomposition terms for C2RMAX10 is extremely similar to the one from 21FR20 (left panel of Fig. \ref{fig:21cmFast_DecFrac} (lower panel)). In the left panel of Fig.~\ref{fig:C2Ray_21star_Bias} we show the absolute value of the bias parameter for C2RMAX10 (red) and C2LLS10 (yellow) for three different stages of reionization, namely $\xb{} \approx 0.7$ (dotted lines), $\xb{} \approx 0.5$ (solid-dash lines) and $\xb{} \approx 0.2$ (thick solid lines). The corresponding BSDs are shown in the right panel. 

The influence of MFP effect is most prominent in the late period $\xb{} \approx 0.2$, seen in the left panel of Fig.~\ref{fig:C2Ray_21star_Bias} (thick solid lines). The choice for the implementation of the MFP effect has a direct influence on the shape of the bias parameters in Fig.~\ref{fig:C2Ray_21star_Bias} (left panel). The bias parameter of C2RMAX10 resembles that of 21FR10 in Fig.~\ref{fig:21cmFASTBias21_xHI_0.2} (left panel), with a similar value of the transition scale following the $k_{\mathrm{trans}} \approx 2/\lambda_{\mathrm{MFP}}$ relation. However we note a more gradual transition than observed in {\sc 21cmFAST}. A real space spherical barrier, therefore, does not yield as sharp of a transition as a $k$-space barrier applied in {\sc 21cmFAST}. 

The bias parameter of the C2LLS10 model exhibits a different behaviour. The implementation imposes a softer limit for the ionizing photons, allowing them to travel distances larger than 10~Mpc. Hence, the large-scale bias is not only boosted but is seen to transition to scale-independence at larger scales, close to $k_{\mathrm{trans}} \lesssim 0.1~\mathrm{Mpc}^{-1}$. The smooth MFP implementation yields an even smoother transition as it affects scales both smaller and larger than the MFP value. This implies that ionized bubbles larger than the MFP value are allowed to develop and influence the shape of the 21-cm signal. We can see this in the right panel of Fig.~\ref{fig:C2Ray_21star_Bias}, where the method applied in C2RMAX10 (light red solid line) favours the formation of bubble sizes up to the hard barrier MFP, causing in a boost in the amplitude of the bubble size distribution. Conversely, the peak of the BSD of C2LLS10 (yellow solid line) is shifted to larger values.

Differences between the C2RMAX10 and C2LLS10 ionization slices are most pronounced in the later stages of the EoR. This becomes evident by examining the morphology of ionized regions for both simulations in Fig.~\ref{fig:C2Ray_Slice}. For C2RMAX10, the spherical implementation of the MFP effect strongly inhibits the formation of bubbles larger than 10~Mpc. Small-scales features are more numerous, and the large-scale morphology is the result of multiple overlapping bubbles of similar sizes. Contrary to this, the C2LLS10 model allows for larger ionized bubbles to form in the densest regions, shaping large-scale structure features accordingly. 

In conclusion, we find that the method of implementing the MFP effect plays a pivotal role in shaping the large-scale bias. The bias values of both {\sc C$^2$-Ray} and {\sc 21cmFAST} are consistent, signifying a direct relation between the MFP value and the transition feature. Studies using both types of simulations by varying the MFP value indicate that a larger MFP accommodates for the formation of large bubbles within the densest regions of the simulation, pushing the scale-independent regime to lower $k$ values, following a $k_{\mathrm{trans}} \approx 2/\lambda_{\mathrm{mfp}}$ relation. The overall shape of the bias is influenced by the choice of ionizing efficiency and minimal mass, and the way that the MFP effect is implemented. Crucially, for the C2LLS10 model, the transition feature has migrated to lower $k$-scales, reaching scale-independence at scales 2 -- 3 larger than given by the above relation. 

\section{Conclusion and discussion}
\label{Sec:Conclusions}
In this work we have investigated the decomposition of the 21-cm power spectrum in terms of the density and neutral fraction fluctuations, similar to what was done in \citetalias{Lidz2007} but focusing on larger length scales corresponding to $k \lesssim 0.3 ~\mathrm{Mpc^{-1}}$. We do not include the effects of potential spin temperature fluctuations. We have employed and compared a set of numerical and semi-numerical reionization simulations and examined the constituent terms throughout the evolution of the average neutral/ionized fractions. Our conclusions are as follows:

\begin{itemize}
\item [1.] For ionized fractions of a few per cent, the $\Delta^{2}_{\delta_{\rho},\delta_{\rho}}$ term dominates and $\Delta^{2}_{\mathrm{21cm}}$ is a scaled copy of the matter power spectrum on large (low-$k$) scales. 

\item [2.] As the other terms gain importance, reionization enters the period of equilibration, characterized by very low amplitudes of the large-scale modes of $\Delta^{2}_{\mathrm{21cm}}$. This period corresponds to ionized fractions of approximately 5 -- 10 per cent. All of the decomposition terms are of comparable strength and the shape of the 21-cm power spectrum neither follows the matter nor the neutral fraction power spectrum, nor indeed any combination of these. This behaviour is caused by the removal of the density peaks from the 21-cm signal due to ionization. As this process is sensitive to the strength of the cross-correlation between ionization and density, the timing and strength of the reduction in amplitude depend on the source parameters. 

\item [3.] After equilibration and until the end of reionization the large-scale $\Delta^{2}_{\mathrm{21cm}}$ becomes a scaled copy of the neutral fraction power spectrum. $\Delta^{2}_{\delta_{\HI},\delta_{\HI}}$ is also  the dominant term, except initially when the cross term $\Delta^{2}_{\delta_{\rho},\delta_{\HI}}$ briefly dominates. The neutral term, and hence the 21-cm power spectrum, acquires a non-monotonic shape, associated with the characteristic bubble scale. 

\item [4.] On the large (low-$k$) scales we have considered the higher order terms are only important during equilibration and during the later stages of the EoR. Specifically, 
$\Delta^{2}_{\delta_{\HI} \delta_{\rho},\delta_{\HI}}$ overtakes $\Delta^{2}_{\delta_{\rho},\delta_{\HI}}$ to become the second most important term after the neutral term beyond ionized fractions of 70 -- 80 per cent.

\end{itemize}

The large-scale 21-cm power spectrum is a biased version of the density power spectrum. However, the growth of regions can mask out information of the density field and profoundly restrict the scales where this assumption is valid. We investigate at which scales this assumption remains true by constructing a bias parameter. We conclude the following:
\begin{enumerate}

\item [5.] Following points 1 and 2, we recover a full scale-independent positive bias early on. The bias becomes strongly scale-dependent and diminishes during equilibration and experiences a sign flip as the 21-cm field becomes anti-correlated with the density field.

\item [6.] After equilibration the bias acquires a two-regime structure, with a scale-independent bias at large-scales and a strongly scale-dependent bias at smaller scales.

\item [7.] The transition point between these two regimes evolves to larger length scales as the position of the peak of the bubble size distribution increases. However, it stops evolving once the scale of the MFP value is reached, following the relation $k_{\mathrm{trans}} \approx 2/ \lambda_{\mathrm{MFP}}$.

\item [8.] At length scales larger than the MFP the 21-cm power spectrum thus remains a biased version of the matter power spectrum. The value of this bias correlates with the source efficiency parameter but does not depend on the minimal mass parameter. During reionization this bias increases as the ionized fraction increases.

\item [9.] While the shape and amplitude of the bias are consistent between {\sc C$^2$-Ray} and {\sc 21cmFAST} results, different implementations of the MFP effect impact the shape of the transition. The $k$-space barrier used in {\sc 21cmFAST} produces a sharp transition, whereas a real space barrier as used in {\sc C$^2$-Ray} produces a smoother transition. Implementing the MFP effect as an absorption process, produces an even smoother transition.

\end{enumerate}
The MFP effect thus plays a crucial role in shaping the two-regime form of the 21-cm power spectrum, as scales larger than the MFP value remain virtually unaffected by astrophysical phenomena and in the most optimistic case could be utilized as an independent probe of cosmology. If not competitive with other such probes, we could at least use
the fact that the slope of the large scale 21-cm power spectrum should tend to the slope of the matter power spectrum as a check on this challenging measurement.

The 21-cm power spectrum has been studied in many previous works and one might wonder whether the above behaviour has not been noted before. Definitely the early works of \citet{Furlanetto2004, Furlanetto2006, McQuinn2006} already highlighted the break feature in the 21-cm power spectra and its relation to a characteristic bubble scale \citep[see fig.~9 in][]{Furlanetto2004}. The sharp break feature we describe in Sec.~\ref{sec:mfp_and_bias} was noted by \citet{Zahn2011}, linking it to the filtering applied within the excursion set principle \citep[see e.g the appendix of][]{Zahn2007}, which is how the MFP effect is implemented in {\sc 21cmFAST}. 
However, the relatively small volumes in many early studies made it hard to recognize the impact of the MFP effect. \citet{Pober2014} studied the impact of the $R_{\mathrm{max}}$ parameter from {\sc 21cmFAST} on the shape of the 21-cm power spectrum in a 400 Mpc volume. They noted that the value of $R_{\mathrm{max}}$ has a strong impact on the large-scale modes and speculated that the position of the break (or knee) could be used to say something about the value of the MFP. However, they did not appear to have noted that the slope of $\Delta^{2}_{\mathrm{21cm}}$ on scales larger than the knee is that of the matter power spectrum. Note that the sharp transition seen at $k \approx 0.07~\mathrm{Mpc}^{-1}$ in the middle panel of their fig.~4, where $R_{\mathrm{max}} = 30$~Mpc, confirms our scaling relation as $k_{\mathrm{trans}} = 2/30 \approx 0.07 ~\mathrm{Mpc}^{-1}$. In fact, once you know what to look for, the break feature due to the $R_{\mathrm{max}}$ value can be seen in many published {\sc 21cmFAST} results, for example in fig.~1 of \citet{Gorce2020} and fig.~4 in \citet{Munoz2021}. 

However, with the few exceptions listed above, the impact of the MFP effect on the shape of the large-scale 21-cm power spectrum seems to have remained underexposed. As we have shown this behaviour becomes most obvious when studying the 21-cm bias. The most extensive investigation of the this quantity has been by \citet{Xu2019}. Their results are consistent with ours up to $\xb \approx 0.2 $, after which the large-scale bias parameter does not display the same scale-independent behaviour we observe in our models, although it does flatten, see their Fig.~10 (a). There are several key differences between their work and ours. For these late stages of reionization, they use the code {\sc islandFAST} \citep{Xu2017} instead of {\sc 21cmFAST} and their MFP value is not constant but has a redshift dependence according to the fit from \citet{Songaila2010}. Furthermore, they included a contribution from \Hi~in galaxies in their bias parameter. The exact cause of the difference between their and our bias results warrants further investigation, which is however beyond the scope of this paper.

The appearance of a sharp break in the {\sc 21cmFAST} results is of course due to the use of a sharp spherical filter in $k$-space. It should be noted that {\sc 21cmFAST} also allows a position- and time-dependent calculation of the MFP value according to the work of \citet{Sobacchi2014} and we expect the break to be much smoother in that case. As we saw from the results in Section~\ref{sec:mfp_impact}, implementing the MFP effect more physically correct as an additional absorption process also softens the transition feature. The observed 21-cm power spectrum is therefore unlikely to have a break as sharp as seen in the constant $R_{\mathrm{max}}$ implementations of {\sc 21cmFAST}. This should be kept in mind when using this version of the code.

All simulations in our study used a constant comoving value for the MFP value. However, as the MFP in the post-reionization Universe is known to have a strong redshift-dependence \citep{Worseck2014}, it would be surprising if it attained a constant value during reionization. The impact of such a redshift-dependent MFP on our results will be interesting to explore but will likely also smooth the transition feature as it will be moving from smaller to larger scales as reionization progresses. Our study has also only explored a single uniform value of the MFP, which in reality varies substantially between different sight-lines \citep[see][for such a model]{2016MNRAS.458..135S} which may further affect the transition towards the linear bias regime.

Ultimately the MFP effect should of course not be implemented as a subgrid model but be modelled self-consistently in a fully hydrodynamic reionization calculation. However, this is at the moment computationally not feasible for reionization simulations spanning many hundreds of Mpc.
However, works such as \citet{Park2016,D'Aloisio2020}, the {\sc THESAN} project \citep{2021arXiv211001628G,Kannan2021,Smith2021} and {\sc CoDA~III} \citep{2022arXiv220205869L} are starting to explore the complex behaviour of the small scale absorbers and can guide us in the future implementation of subgrid models.

Lastly, we would like to discuss whether the effects we found can actually be detected in observations. The largest length scales are more likely to be affected by the residuals from smooth foreground signals and so some of the values considered in this paper will remain unattainable for the present and upcoming telescopes. However, at present it is difficult to know the lowest $k$ which can be reliable extracted. For example \citet{2021arXiv211003190R} arrive at a minimal value of approximately 0.1~Mpc$^{-1}$ for data from the MWA whereas \citet{Mertens2020} manage to subtract the foreground signals down to $k = 0.0525~\mathrm{Mpc}^{-1}$. While it remains true that the shape of $\Delta^{2}_{\mathrm{21cm}}$ for $k\lesssim 0.3$~Mpc$^{-1}$ is the shape of the neutral fraction power spectrum for most of reionization, the transition to a scale-independent bias may or may not be detectable, depending on the effectiveness of the foreground mitigation techniques and of course the scale at which this transition occurs. The recent measurements of the mean free path at $z = 6$ by \citet{Becker2021} would imply that the transition will be found in the range $k_{\mathrm{trans}}=0.2$ -- 0.9 $\mathrm{Mpc}^{-1}$ at that redshift. Furthermore, as the MFP is likely to be shorter at larger redshifts, even a transition around 0.05~Mpc$^{-1}$ at $z=6$ would imply $k_{\mathrm{trans}}\approx 0.25  ~\mathrm{Mpc}^{-1}$ at $z=8$ according to the redshift scaling from \cite{Worseck2014}. We therefore conclude that the transition to the cosmological regime may be measurable during parts of reionization even if foreground mitigation is ineffective for $k< 0.1~\mathrm{Mpc}^{-1}$. Estimating the bias from future measurements of the 21-cm power spectrum could thus allow us to put observational constraints on the value of the MFP during the EoR.

Given the impact of the MFP implementation on the large-scale 21-cm power spectrum, future simulation efforts should carefully consider not only the value but also the implementation of the MFP effect when using their results, for example in the context of parameter estimation. The MFP value has often been seen as the least important of the basic parameters of the reionization process. Our results show that it in fact has substantial impact on the reionization process and the resulting 21-cm power spectrum. This therefore warrants improvements in the implementation of the MFP effect in both numerical and semi-numerical simulations of reionization. 

In terms of future work, we plan to further investigate the connection of our large-scale bias values to proposed linear bias prescriptions such as the one in \citet[][see eq.~6.1]{McQuinn2018}. This would also make the connection to the halo bias of the source population. Moreover, we plan to investigate the effect of using a more physically motivated MFP implementation in which the subgrid absorptions are both spatially and redshift dependent, along similar lines as the implementation in \citet{2016MNRAS.458..135S}. Such simulations should make it clearer what can be expected from measurements of the real large-scale 21-cm power spectrum.

\section*{Acknowledgements}
This work was partly inspired by earlier work of Sander Bus and Saleem Zaroubi on the decomposition of the 21-cm power spectrum which remained unpublished. We also wish to thank Raghunath Ghara for his assistance and useful discussions and Andrei Mesinger for providing clarifications about {\sc 21cmFAST}. We thank the anonymous referee for helpful and constructive comments. GM acknowledges support by Swedish Research Council grant 2020-04691. RM is supported by the Wenner-Gren Postdoctoral Fellowship.
Some of the numerical results were obtained via supercomputing time awarded by PRACE grants 2012061089, 2014102339, 2014102281 and 2015122822. Other results were obtained on resources provided by the
Swedish National Infrastructure for Computing (SNIC) at
PDC, Royal Institute of Technology, Stockholm.

We have utilized the following python packages for manipulating the simulation outputs and plotting results: numpy \citep{harris2020array}, scipy \citep{2020SciPy-NMeth}, matplotlib \citep{Hunter2007} and{ \sc tools21cm}\footnote{\url{https://github.com/sambit-giri/tools21cm}} package \citep{Giri2020}.

\section*{Data Availability}
The data underlying this article will be shared on a reasonable request
to the corresponding author.



\bibliographystyle{mnras}
\bibliography{example} 




\appendix
\section{Decomposition of a reionization model with rare and bright sources. }
\label{appendix:C2RBL70}

\begin{table}{h}
\centering
\caption{{\sc C$^2$-Ray} simulation source parameters. For this simulation, the number of ionizing photons escaping from a halo every $10^7$ years is given by $g_\gamma$ times the number of baryons in the halo. This source model has been used in most previous {\sc C$^2$-Ray} results \citep[see e.g.][]{2016MNRAS.456.3011D,giri2019position}. } 
\label{tab:C2RBL70_sim}
\begin{tabular}{c|c|c|c|c} 

Label & \multicolumn{1}{p{1cm}|}{Source \newline Model}  & \multicolumn{1}{p{1cm}|}{$M_{\mathrm{min}}\newline [M_{\odot}]$} & \multicolumn{1}{p{1cm}|}{MFP \newline Method} & \multicolumn{1}{p{1cm}|}{$\lambda_{\mathrm{MFP}} \newline $[Mpc]} \\ \hline
C2RB$\lambda$70  & $g_\gamma=2550$ & $5 \times 10^{10}$ & cubical & 70\\

\end{tabular}
\end{table}

\begin{figure}
\includegraphics[width=\columnwidth]{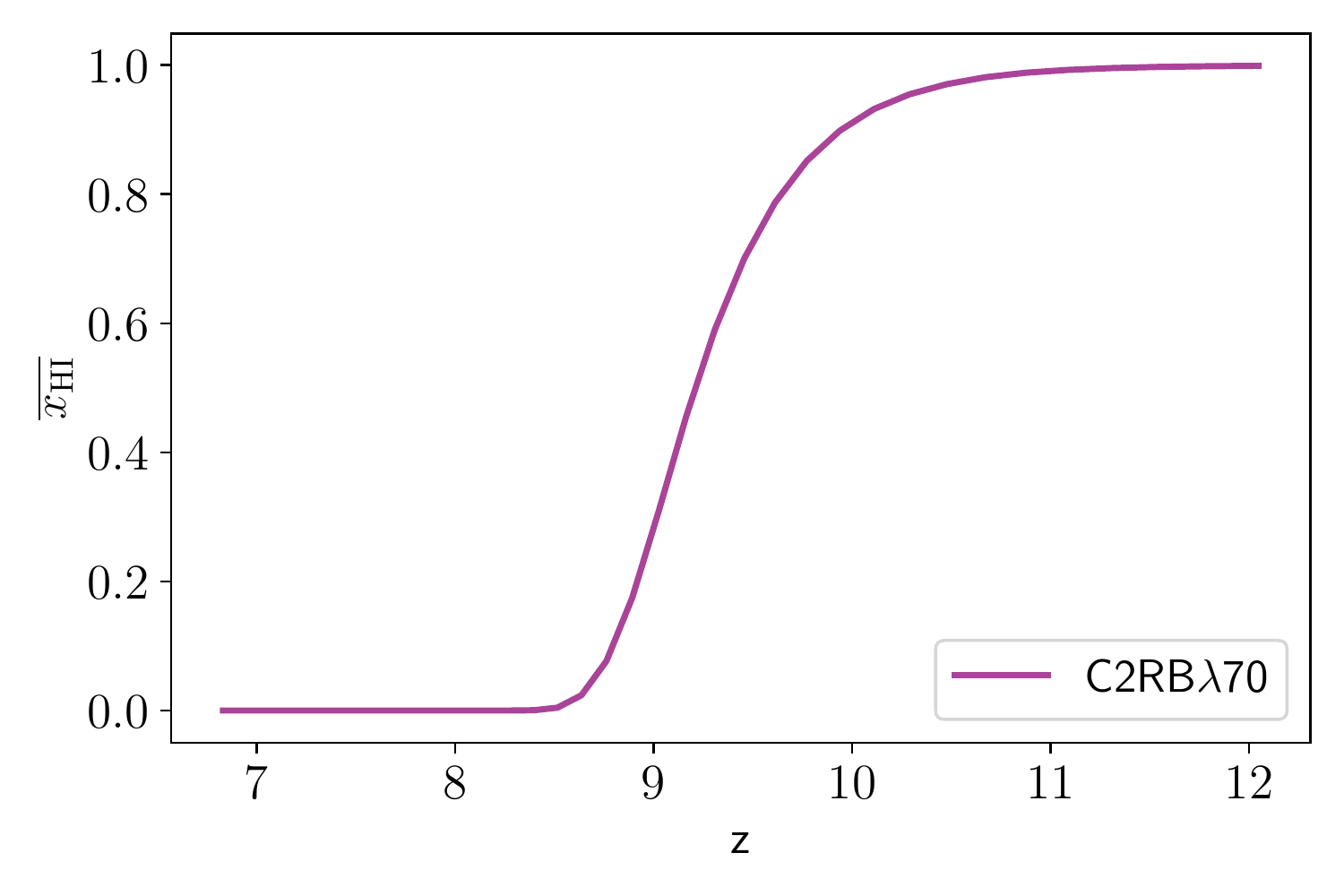}
\includegraphics[width=\columnwidth]{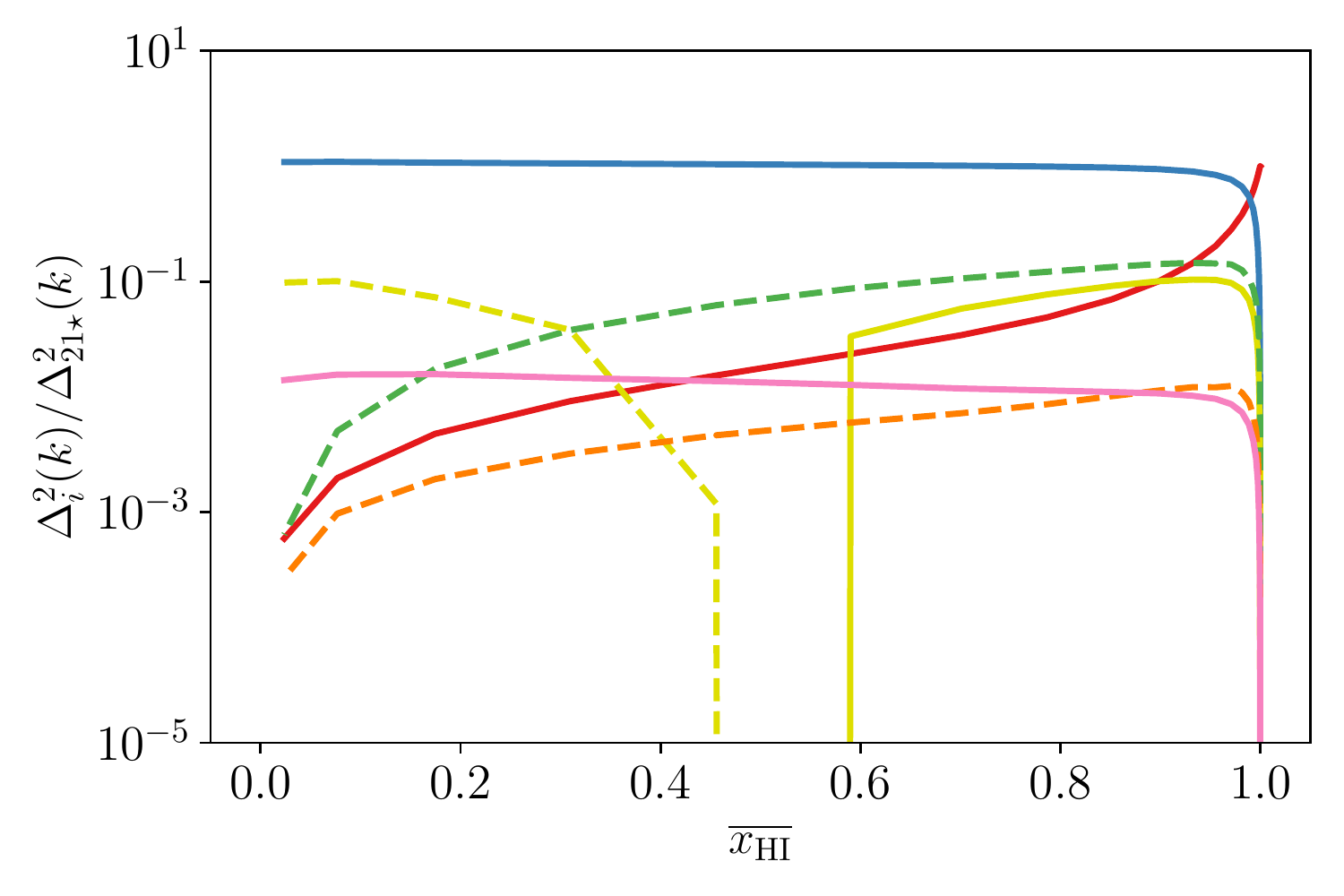}
\caption{Upper Panel: Reionization history of the C2RB$\lambda$70 model in purple.
Lower Panel: Fractional contribution of each constituent term in Eq. \ref{Eq::Decomposition} at $k = 0.1~\mathrm{Mpc}^{-1}$ throughout the EoR. Solid lines mark positive terms, dashed lines represent the absolute value of the negative terms of the C2RB$\lambda$70 model.}
\label{fig:C2RBL70_DecFrac}
\end{figure}

In this appendix we show the evolution of the decomposition terms for an additional {\sc C$^2$-Ray} simulation in which reionization is driven by relatively few but very bright sources. For such a source model the correlation between density and ionization is severely weakened. The astrophysical parameters for this simulation are listed in Tab.~\ref{tab:C2RBL70_sim} and the reionization history in provided in Fig.~\ref{fig:C2RBL70_DecFrac} (upper panel). We should point out that this simulation is unlikely to be consistent with any of the existing observational constraints on reionization and was in fact developed as an example of a scenario that would be ruled out by the LOFAR $z=9.1$ upper limits on the 21-cm power spectrum \citep{ghara2020constraining}. We include it here as an illustration of how a weakened correlation between density and ionization can affect the decomposition.

The lower panel of Fig.~\ref{fig:C2RBL70_DecFrac} shows the evolution of relative contribution of the decomposition terms against the average neutral fraction and can be compared with Fig.~\ref{fig:21cmFast_DecTerms}. We see that for this scenario, the whole period of equilibration disappears. The 21-cm power spectra directly transitions from being dominated by the density term to being dominated by the neutral hydrogen term. The contribution of the cross-term (green line) never reaches more $\sim 0.1$. The extreme efficiencies of the sources lead to every source ionizing a large volume around itself, which thus contains both low and high density areas, weakening the correlation between ionization and density. For this weak correlation the required conditions to launch equilibration are suppressed. Note however, that even for this extreme case, the evolution of the decomposition terms after equilibration is qualitatively similar to what we saw before, further attesting to the robustness of the decomposition.

\section{Implementations of the mean free path in simulations}
\label{appendix:MFP_application}
In this paper we use results from two different codes, namely {\sc 21cmFAST} and {\sc C$^2$-Ray}. The former is a so-called semi-numerical code developed from the idea of using an excursion set method to determine whether a region is ionized or not \citep{Furlanetto2004}. The latter is a fully numerical ray-tracing code \citep{Mellema2006}. Due to this fundamental difference, the effect of a finite MFP for ionizing photons in the ionized IGM are implemented differently.

In {\sc 21cmFAST} the excursion set method considers different scales when checking whether cells should be ionized or not. It starts these checks at a maximum scale $R_\mathrm{max}$ and as shown by \citet{Davies2021} this corresponds to a hard horizon beyond which sources cannot influence a given location. As these calculations are performed in the Fourier domain, the impact of $R_\mathrm{max}$ on a Fourier domain quantity such as the power spectrum can be expected to be rather sharp, as we have seen.

Newer versions of {\sc 21cmFAST} implement the method described by \citet{Sobacchi2014} in which the MFP becomes both redshift and position-dependent. We expect that the use of this approach will make the transition between the scale-dependent and scale-independent bias regimes smoother as it will introduce a range of $R_\mathrm{max}$ values for every redshift.

However, as \citet{Davies2021} reminded us, the use of a hard barrier at a fixed radial distance is a rather crude implementation of the effect of the MFP. Rather the absorption of ionizing photons in the ionized IGM is a stochastic process which is better described through an optical depth where after a given distance $\lambda_\mathrm{MFP}$ an optical depth of 1 is reached. Those authors implemented a modification in {\sc 21cmFAST} which uses this gradual absorption approach. Also in this case we would expect the transition in the bias to be more gradual.

{\sc C$^2$-Ray}, being a ray-tracing radiative transfer code, offers several different implementations of the MFP effect. The simplest is again a hard barrier, implemented as the largest distance for which to follow the radiation from a given source. Due to the nature of the ray-tracing on a Cartesian grid, this barrier is most easily be implemented as having a cubic shape, however, the code also offers the option of using a (discretized) spherical barrier. For larger values of the MFP the difference between the two shapes has little effect on the outcome. Most published {\sc C$^2$-Ray} results used the cubic shape (typically with $\lambda_\mathrm{mfp}=70$~Mpc) but the hard barrier simulation used in this paper, C2RMAX10, uses the spherical barrier.

\citet{2016MNRAS.458..135S} already developed a MFP model using gradual absorption for {\sc C$^2$-Ray}, similar to the approach promoted by \citet{Davies2021} for semi-numerical excursion set methods. In this model additional absorption is applied before passing photons from one cell to the next, making sure that the associated opacity reaches an optical depth of 1 for Lyman limit photons at the required MFP distance. Note that compared to the semi-numerical case, a fully numerical implementation includes the frequency-dependence of the absorption process. In \citet{2016MNRAS.458..135S} the MFP distance was taken to be redshift-dependent, for example following the evolution of the MFP suggested by \citet{Songaila2010}. The same authors also explored the effect of a position-dependent opacity, adding more opacity to cells that contain more mass in dark matter haloes. For this paper we opted for the simpler constant MFP value of 10~Mpc and a uniform distribution of the associated opacity.

\bsp	
\label{lastpage}
\end{document}